# Internal sensory models allow for balance control using muscle spindle acceleration feedback

Eric Maris[*], [#]

[*]*Donders Institute for Brain, Cognition, and Behaviour, Radboud University, P.O. Box 9104 HE Nijmegen*

[#]

Corresponding Author:

Eric Maris

e-mail: eric.maris@donders.ru.nl

tel: +31243612651

The number of words of the abstract: **225**

The number of figures: **4**

The number of tables: 0

## Abstract

Motor control requires sensory feedback, and the nature of this feedback has implications for the tasks of the central nervous system (CNS): for an approximately linear mechanical system (e.g., a freely standing person, a rider on a bicycle), if the sensory feedback does not contain the state variables (i.e., joint position and velocity), then optimal control actions are based on an internal dynamical system that estimates these states from the available incomplete sensory feedback. Such a computational system can be implemented as a recurrent neural network (RNN), and it uses a sensory model to update the state estimates. This is highly relevant for muscle spindle primary afferents whose firing rates scale with acceleration: if fusimotor and skeletomotor control are perfectly coordinated, these firing rates scale with the exafferent joint acceleration component, and in the absence of fusimotor control, they scale with the total joint acceleration (exafferent plus reafferent). For both scenarios, a sensory model exists that expresses the exafferent joint acceleration as a function of the state variables, and for the second scenario, a sensory model exists that corrects for the reafferent joint acceleration. Simulations of standing and bicycle balance control under realistic conditions show that joint acceleration feedback is sufficient for balance control, but only if the reafferent acceleration component is either absent from the feedback or is corrected for in the computational system.

# Introduction

Effective goal-directed movements almost always require sensory feedback. For example, standing upright depends on sensory feedback about the body's center of gravity (CoG) relative to its area of support (AoS). The central nervous system (CNS) uses this feedback to compute control actions (ankle torques) that overcome the toppling torque due to gravity and prevent the body from falling. This is similar in other forms of balance control, such as cycling: sensory feedback now pertains to the combined CoG of rider and bicycle relative to the line of support (LoS, the line that is formed by connecting the two tire-road contact points), and the control actions involve turning the handlebars.

The difficulty of the CNS's control task depends on the sensory feedback. It is useful to investigate this point from a control theoretic perspective. I will do this for a musculo-skeletal system (a human body plus attached objects like a bicycle) that can be approximated by a linear mechanical system. A mechanical system is linear if the limb movements are described by differential equations (equations of motion; EoM) that are linear in the joints' angular positions and velocities, which are the system's state variables. For balance control, the mechanical system is linear near the so-called fixed point, and for a standing body and a rider-bicycle combination, this fixed point is the upright position (CoG above the AoS/LoS). Therefore, because most of the movements remain close to the fixed point, a linear approximation of the nonlinear EoM captures these movements well. Now, the optimal control actions for a linear mechanical system are a simple linear combination of the state variables [1]. Thus, the control task for the CNS is simple if the sensory feedback is an invertible linear function of the state variables (denoted as full state feedback):

compute optimal control actions as a simple linear combination of this full state sensory feedback.

Under normal conditions, the sensory feedback that is used for balance control involves multiple sensory modalities: the contributions of somatosensory (including proprioceptive), vestibular, visual and auditory feedback have all been identified [2]. Here, I will focus on proprioceptive feedback from the muscle spindles [3]. There is good empirical evidence that, at least on a firm and immobile support surface, muscle spindle proprioceptive feedback is sufficient for standing balance control [4-6]. Whether the muscle spindles provide feedback about angular position and velocity (i.e., full state feedback) is a topic of much debate in sensory neurophysiology [7-9]. I will put this debate in a different perspective by (1) only considering a type of sensory feedback about which there is no controversy, namely muscle spindle primary afferents, whose firing rates encode acceleration [10-13], and (2) demonstrating that the CNS can estimate the state variables from acceleration feedback. Thus, I will shift the focus from the properties of the sensory organs to computations at the level of the CNS.

I will describe two methods with which the CNS can estimate the state variables from acceleration feedback, and both involve an internal sensory model. The first model builds on the fact that, under perfectly coordinated skeletomotor and fusimotor control, the firing rates of muscle spindle primary afferents scale with the exafferent joint acceleration component. The second model assumes no fusimotor control, and in this case the primary afferent firing rates scale with the total (exafferent plus reafferent) joint acceleration.

Each of these sensory models can be combined with the familiar internal model for the dynamics of the mechanical system [14, 15], and together they specify a dynamical system that estimates the current state. This CNS-internal dynamical system can be implemented as a recurrent neural network (RNN). Crucially, both methods for estimating state variables from acceleration feedback prevent reafferent feedback from entering the CNS as noise and disrupting computations.

The simulation methodology used in this paper allows to evaluate the destabilizing consequences of ignoring reafferent feedback. Specifically, I have simulated balance control with or without a mechanism that corrects CNS-level computations for the reafferent feedback. This allows to evaluate whether successful balance control depends on such a mechanism.

The evidence in the present paper comes from formal analysis and computer simulations, and the main point can be introduced using the one degree-of-freedom (1-DoF) inverted pendulum model of standing balance control. Focusing on such a simple model has the risk of making claims that are not valid for more complicated systems. I will therefore make the same point using two 3-DoF models for bicycle balance control. Balancing a bicycle using only proprioceptive feedback is a very challenging control task because the mechanical system is underactuated [16]: the rider is not touching the road surface, and therefore the bicycle lean angle cannot be directly registered by the muscle spindles nor controlled by a forcing torque over this mechanical joint. I will demonstrate that a bicycle can be balanced using only acceleration feedback from the rider's upper body and a steering torque that controls the front frame via the handlebars. Crucially, this only holds if the reafferent acceleration component is either cancelled at the source (i.e., the muscle spindle) or is corrected for at the level of the CNS.

All simulations were run using biologically realistic parameter values, except for the sensorimotor delay (SMD; the time between the acceleration reflected in the sensory feedback and the effect of the control action) which was set at 0, whereas realistic SMD values have been reported in the range between 100 and 160 ms [2, 17-19]. The simulation results for bicycle balance control strongly depend on the presence of a nonzero SMD: for SMDs as small as 20 ms, rider and bicycle touch the road surface [20]. However, powerful and neurobiologically plausible SMD compensation mechanisms exist that allow for balance control of the rider-bicycle combination with an SMD of 560 ms and, crucially, while using exafferent acceleration feedback [20]. The description, motivation and evaluation of these SMD compensation mechanisms (i.e., memory-informed state estimation and prediction) are described in a different paper [20].

# Results

### A closed-loop feedback control model for standing balance

In Fig. A and 1B, a closed-loop feedback control model is shown [21-23]. In its application to standing balance control, the mechanical system (in red) is the body's musculo-skeletal system, which is depicted separately in Fig. 1C. This mechanical system is modeled as a compound inverted pendulum (CIP) that rotates about the ankle joint. The pendulum's angular position relative to gravity (lean angle) is denoted by $\theta$ and the muscular forcing torque at the ankle joint by $\boldsymbol{z}$. The movements of the CIP are fully specified by its equation of motion (EoM), which expresses the acceleration $\ddot{\theta}$ as a function of position $\theta$, velocity $\dot{\theta}$, and the forcing torque $\boldsymbol{z}$. The EoM is usually written as a function of the mechanical system's state, which is the combined position $\theta$ and velocity $\dot{\theta}$, and is denoted by $\boldsymbol{x} = [\theta, \dot{\theta}]^t$. The distinction between the state $\boldsymbol{x}$ and the acceleration $\ddot{\theta}$ is crucial for balance control because the muscular forcing torque $\boldsymbol{z}$ controls $\ddot{\theta}$ in a direct way, whereas the state variables $\theta$ and $\dot{\theta}$ are only controlled indirectly via $\ddot{\theta}$. The EoM can be written as a differential equation for the state $\boldsymbol{x}$, which I will denote as $\dot{\boldsymbol{x}} = \Phi(\boldsymbol{x}, z)$. This differential equation is usually called the EoM in state-space form, and it is described in more detail in *An optimal computational system to evaluate the sufficiency of sensory feedback*. (Here and in the following, I will not follow the notational convention of using lower case boldface letters for vectors and normal font for scalars; all results in this paper hold for the vector-valued case, but sometimes the application/example involves only scalars.)

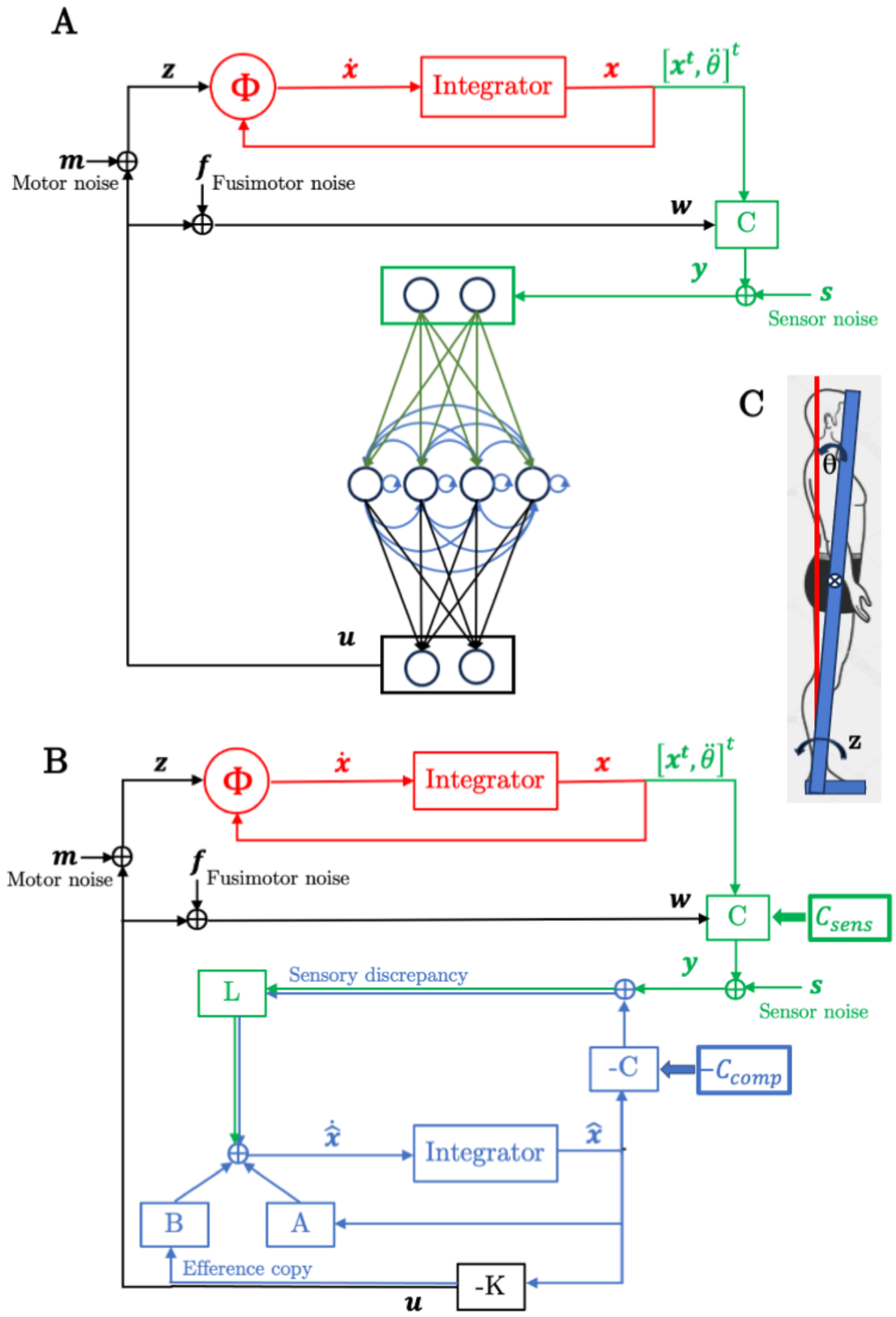

**Fig. 1: Closed-loop feedback control model for standing balance.** (A) Block diagram of the closed-loop feedback control model, with a mechanical (in red), a sensory (in green), a computational (multiple colors; see below), and a motor output system (in black). The computational system is a neural network with an input layer with input $\boldsymbol{y}$, an intermediate layer with recurrent connections, and an output layer with output $\boldsymbol{u}$. (B) The computational system is based on a linear internal model (approximation) $\dot{\boldsymbol{x}} = A\boldsymbol{x} + B\boldsymbol{z}$ of the nonlinear mechanical system $\dot{\boldsymbol{x}} = \Phi(\boldsymbol{x}, \boldsymbol{z})$ and a sensory internal model $C$. This assumption, together with an optimality criterion for the control performance (see Eq. 1), defines the weights of the neural network. The matrices $L$ and -$K$ are, resp., the Kalman and the LQR gain (see text). (C) A standing body with a superimposed compound inverted pendulum rotating about the ankle,

used to model standing balance control. The pendulum's angular position relative to gravity is denoted by $\theta$ and the torque at the ankle joint by $\boldsymbol{z}$.

The sensory system (in green; a model of the sensory nervous system) linearly maps the vectors $[\boldsymbol{x}^{\boldsymbol{t}}, \ddot{\theta}]^t$ and $\boldsymbol{w}$ onto sensory variables, adds noise $\boldsymbol{s}$ and feeds the resulting $\boldsymbol{y}$ into the computational system (in blue; a model of the CNS). With respect to the sensory variables, I deviate from the existing literature on closed-loop feedback control, where the sensory feedback pertains to the state variables $\boldsymbol{x}$. The angular acceleration $\ddot{\theta}$ refers to the total joint acceleration, and the torque variable $\boldsymbol{w}$ is used to correct this total acceleration for its reafferent component. Importantly, the torque variable $\boldsymbol{w}$ contains direct input from the computational (in blue) into the sensory system (in green). In the muscle spindles, this direct input is implemented by spinal gamma and beta motor neurons (fusimotor neurons) and is denoted as fusimotor control [3, 7, 8]. The motor output system (in black) adds noise to the computational system's output $\boldsymbol{u}$, separately for each of the two pathways: (1) adding motor noise $\boldsymbol{m}$ (skeletomotor noise plus external perturbations; see further) produces the noisy torque input $\boldsymbol{z}$ into the mechanical system, and (2) adding fusimotor noise $\boldsymbol{f}$ produces the noisy torque input $\boldsymbol{w}$ into the sensory system.

From a neurobiological perspective, the computational system is a neural network with an input layer, an intermediate layer with recurrent connections, and an output layer (see Fig. 1A). These recurrent connections agree with the fact that, within the CNS, feedback connections are ubiquitous. As is common in neuroscience, I assume that the CNS learns optimal weights for the neural network's connections [21-23]. Calculating these optimal weights is a computational challenge which often is considered a part of machine learning. Here, I take a different approach: I establish a correspondence between the neural network and an equivalent algebraic approach for which the optimal solution is known. This algebraic approach is known as optimal feedback control (OFC), and the optimal solution computes (1) an optimal internal state estimate $\hat{\boldsymbol{x}}$ by integrating a linear differential equation that takes as input an efference copy $\boldsymbol{u}$ and sensory feedback $\boldsymbol{y}$, and (2) an optimal control action $\boldsymbol{u}$ by multiplying the state estimate $\hat{\boldsymbol{x}}$ by the gain -*K*.

The computational system is based on (1) a linear forward model $\dot{\boldsymbol{x}} = A\boldsymbol{x} + B\boldsymbol{z}$ that approximates the nonlinear mechanical system $\dot{\boldsymbol{x}} = \Phi(\boldsymbol{x}, \boldsymbol{z})$, and (2) a linear sensory model $\boldsymbol{y} = C\boldsymbol{x} + \boldsymbol{s}$ of the input to the computational system. I assume the matrices *A*, *B* and *C* to be known/learned and ask for which values of *K* and *L* some optimality criterion is achieved. In the time domain, optimality is usually expressed in terms of a loss function, and in the Laplace domain, it is expressed in terms of the poles of a transfer function.

For the claims of the present paper, the exact mathematical specification of the optimality criterion is not required, but for the simulations it is. For these simulations, I will use linear-quadratic-Gaussian (LQG) control [1]. Applying the LQG results, the optimal input weights for the neural network in Fig. 1A are the Kalman gain *L*, and the optimal output weights are the linear quadratic regulator gain $-K$. From these results, it follows that the optimal recurrent weights are $A - BK - LC$. Of course, the CNS can only learn these optimal weights from experience. Because this process develops over time, the weights are never optimal but converge to optimality. To prevent possible misunderstandings, the equivalence between the neural network in Fig. 1A and the algebraic formalism in Fig. 1B is only used to demonstrate its neural plausibility; no neural network implementation with nonlinear properties has been built.

The LQG is for continuous signals only, whereas the signals in the peripheral, central and somatic nervous system are sets of spike signals. However, it is possible to formulate a spiking neural network (SNN) with the same functional properties as a LQG controller [24]. This functional equivalence can be demonstrated using a SNN [25, 26] that produces the type of spiking activity that is observed in biological neural networks: irregular, sparse, and robust.

**Incomplete feedback**

For a linear mechanical system with observable states $\boldsymbol{x}$, the optimal control action is the linear combination $-K\boldsymbol{x}$ [1]. Optimal control actions can also be computed from the sensory feedback when the sensory system's output matrix $C$ is invertible, namely as $-KC^{-1}\boldsymbol{y}$. This is full state feedback, and there is no need to pass such feedback through a recurrently connected intermediate layer such as in Fig. 1A. (To prevent possible confusion, note that "full state feedback" is also used in a different meaning, namely as a method for pole placement in classical control theory.)

If the sensory feedback is not full state, it will be called incomplete. Incomplete feedback can be of different types, such as a non-invertible mapping of the mechanical system's state $\boldsymbol{x}$, or a mapping of non-state variables such as acceleration $\ddot{\theta}$ and torque $\boldsymbol{w}$ (see Fig. 1A and 1B). In this paper, I will only focus on the latter type. However, for all types of incomplete feedback, it holds that it must be passed through a recurrently connected intermediate layer if it is to be converted into an optimal control action.

Incomplete feedback is not the only use case for a recurrently connected intermediate layer. Such a layer will also reduce the effects of motor and/or sensory noise that is added to the input and/or output of the mechanical system. If the sensory feedback is noisy, more accurate state estimates are obtained by combining the noisy sensory feedback with model-based predictions using Bayes' rule [27]. Crucially, this combination can be performed by the RNN in Fig. 1A, and the optimal combination of the noisy sensory feedback with the model-based predictions is realized by the Kalman gain $L$. The difference between noisy full state and incomplete feedback is that the RNN does not simply improve the accuracy of the state estimates: it bridges the gap between incomplete feedback and full state information.

## Muscle spindle proprioceptive feedback

### *Proprioceptive feedback for balance control*

Balance control relies on contributions from multiple sensory modalities: somatosensory (which includes proprioception), visual, vestibular and auditory [2, 12, 28, 29]. In this paper, I focus on proprioceptive feedback and muscle spindles are claimed to be the most important sensory organ for this modality [3]. I will formulate a model for balance control that uses only muscle spindle proprioceptive feedback. This is a plausible hypothesis, at least when the support surface is firm and immobile; on such a surface, humans can maintain standing balance after removing vestibular, touch, and visual feedback [4-6].

### *What feedback is generated by muscle spindles?*

The nature of muscle spindle afferent feedback is a topic of debate in sensory neurophysiology: two specialized review papers argue against the claim that the spindles provide feedback about angular position and velocity [7, 8], and one argues in favor of it [9]. Central to this debate is the existence of two types of spindle afferents: the primary (type Ia) and the secondary (type II) afferents. There is consensus that primary spindle afferents do not reliably encode position and velocity, and the strongest empirical evidence comes from studies in which the spindle-containing muscle has to act against a load [30-35]: primary afferent firing rates scale with load (which is related to acceleration; see further), not with joint angular position. This is highly relevant for balance control, which involves acting against a gravitational load.

Given the consensus on the primary afferents, the debate focuses on whether secondary afferents reliably encode position and velocity. In this paper, I take a different approach: I only consider primary afferents and ask the question whether the control-relevant state variables can be estimated from their output. Older studies, reviewed in [7], typically characterize the primary afferents as having a high dynamic sensitivity, which means that they respond with a burst of action potentials at the onset of the stretch and stop responding when the stretch is continued at a constant velocity. More recent studies have investigated the response properties of the spindle afferents by quantifying the degree to which they encode position, velocity and acceleration information. These studies have demonstrated that primary afferents predominantly encode acceleration [10-13].

Thus, I will deviate from the sensory neurophysiology studies whose aim is to demonstrate that the control-relevant state variables are already present in the output of the sensory organs. Instead, I will argue that the CNS estimates these state variables from acceleration feedback. Such a computational mechanism is especially beneficial when the position amplitude is low. This holds for standing balance control: the body sway/lean angular positions relative to gravity are smaller by an order of magnitude than the differences between the joint angular positions that were used to demonstrate the secondary afferents' position sensitivity [9].

*Muscle spindle anatomy and physiology*

A muscle spindle contains the endings of the sensory neurons, and these terminate at so-called intrafusal fibers, which have a sensory function only; the contractile force is produced by the muscle's extrafusal fibers. The sensory neurons' endings respond to stretch in the intrafusal fibers and if they lie slack (zero tension), the sensory neuron provides no information about the muscle's extrafusal fibers. The primary sensory afferents terminate at the central (equatorial) region of the intrafusal fibers, and the secondary afferents terminate adjacent to this central region.

Response properties of muscle spindles are investigated using direct recordings of their output, afferent spikes in sensory neurons. In anesthetized animals, such recordings are obtained from the dorsal horn of the spinal cord [11], and in awake humans from sensory axons using microneurography [7, 36]. One of the main results of human microneurography pertains to the differences between passive and active joint movements: in a passive joint (i.e., a joint that is moved by the experimenter), spindle afferents encode length and velocity information, but this is no longer the case if this joint is actively moved [7].

A crucial difference between the active and the passive muscle is that, in the former, the muscle spindle intrafusal fibers are activated by spinal fusimotor (gamma and beta) neurons as a part of fusimotor control. Without fusimotor control, muscle spindle output would only depend on the extrafusal fibers that actuate the joint. The simplest form of fusimotor control is alpha-gamma coactivation, which effectively deals with the problem that the stretch-sensitive sensory endings have a limited operating range: when the extrafusal fibers contract, the tension in the intrafusal fibers can become too low (they fall slack) for the sensory endings to fire, regardless of the length/velocity/acceleration of the extrafusal fibers. Fusimotor neurons control the length/velocity/acceleration of the intrafusal fibers, and thereby the sensitivity of the sensory neurons. In alpha-gamma coactivation, the gamma and beta motor neurons fire in sync with the spinal alpha motor (skeletomotor) neurons, whose firing makes the extrafusal fibers contract. This coordination keeps the intrafusal fibers at a length that keeps the sensory neurons in their operating range.

More sophisticated forms of fusimotor control have been described and they all involve CNS-level computations that allow for coordination between fusimotor and skeletomotor neurons [8, 37, 38]. Here, I focus on the role of this coordination in removing the reafferent component from the total (exafferent plus reafferent) acceleration feedback. The better the coordination, the more the sensitivity of the sensory neurons shifts to non-muscular forces such as gravity, elasticity (which both depend on angular position) and damping (which depends on angular velocity). If the coordination is perfect, then the net forces acting on the intrafusal fibers are non-muscular, and the muscle spindle feedback is purely exafferent. Crucially, reafferent acceleration is not only produced by the spindle-bearing muscle, but by all muscles that act on the same joint (e.g., agonist muscle intrafusal fiber acceleration produced by the antagonist muscle). This requires a type of fusimotor control that is more sophisticated than alpha-gamma coactivation: it involves awareness by the CNS of muscle synergies, co-activation patterns of muscles recruited by a neural command.

In the next section, I will describe the evidence in favor of the hypothesis that coordination between fusimotor and skeletomotor control produces exafferent acceleration feedback. I will also describe the scenario of no fusimotor control, in which case total acceleration feedback is produced. Crucially, in both scenario's and using scenario-specific methods, the CNS can extract state information from the acceleration feedback.

## A mechanical model for standing balance constrains the relation between intrafusal fiber acceleration and the state variables

### *A mechanical model specifies acceleration components*

Muscle spindles are mechanoreceptors whose firing rate depends on the intrafusal fiber acceleration. I will use a mechanical model for standing balance (the CIP) to specify the acceleration components to which the spindle primary afferents respond. The CIP EoM is this second-order differential equation:

$$\ddot{\theta} = M^{-1}\left[-c_{damp}\dot{\theta} + \frac{1}{2}mgl\sin(\theta) - k_{stiff}\,\theta + z\right] \qquad \text{Eq. 1}$$

In this equation, $m$ is the mass of the compound pendulum, $l$ is its length, and $g$ is the gravitational constant. The constant $M$ denotes the mass moment of inertia $M = \frac{1}{3}ml^2$; it is denoted by a capital letter to distinguish it from the mass $m$ and for consistency with the matrix form of the mass moment of inertia, which will be introduced in a later section. The ankle joint has both stiffness and damping, and these are characterized by the coefficients $k_{stiff}$ and $c_{damp}$. The variable $z$ is the forcing torque applied at the ankle. This forcing torque is produced by the muscles' extrafusal fibers and is the sum of a control signal plus motor noise.

It is useful to linearize Eq. 1 by making use of the fact that, near $\theta = 0$ (the upright/balanced position), $\sin(\theta) \approx \theta$:

$$\ddot{\theta} \approx M^{-1}\left[-c_{damp}\dot{\theta} + \left(\frac{1}{2}mgl - k_{stiff}\right)\theta + z\right] \qquad \text{Eq. 2}$$

A critical aspect of Eq 2 is the relation between $k_{stiff}$ and the critical stiffness $\frac{1}{2}mgl$: if $k_{stiff} < \frac{1}{2}mgl$, the CIP is unstable in the upright position ($\theta$ accelerates away from 0), and if $k_{stiff} > \frac{1}{2}mgl$, it is stable. Several studies have demonstrated that the ankle joint stiffness $k_{stiff}$ is less than the critical stiffness $\frac{1}{2}mgl$, and that this is due to the compliance of the Achilles tendon [39-42].

Eq. 2 can be simplified by using the coefficients $c = -M^{-1}c_{damp}$ and $k = M^{-1}\left(\frac{1}{2}mgl - k_{stiff}\right)$:

$$\ddot{\theta} \approx c\dot{\theta} + k\theta + M^{-1}z \qquad \text{Eq. 3}$$

The first two terms in the right side of Eq. 3 are the exafferent acceleration components; one term ($k\theta$) depends on gravity and stiffness, and the other ($c\dot{\theta}$) depends on damping. The third term in the right side of Eq. 3 is the reafferent acceleration component. Eq. 3 imposes a constraint on the relation between the variables $\theta$, $\dot{\theta}$, $\ddot{\theta}$ and $z$, and the CNS can use this constraint to estimate the state variables from acceleration feedback.

Eq. 3 pertains to a joint that is controlled by multiple muscles, and these are usually grouped functionally as agonists and antagonists. I define the agonists and antagonists as the muscles that produce positive, resp., negative torque. The forcing torque $z$ in Eq. 3 is the net torque that results from the combined action of agonists and antagonists.

*Correspondence between angular and translational accelerations*

I want to use Eq. 3 to specify the acceleration components that are registered by the spindle primary afferents in the joint-crossing muscles (agonists and antagonists). However, Eq. 3 involves an angular acceleration $\ddot{\theta}$ whereas the spindles respond to translational accelerations in the agonist and antagonist muscles. Fortunately, there is a one-to-one correspondence between the joint's angular acceleration $\ddot{\theta}$ and the translational acceleration in the antagonist and the agonist muscles. I will denote these translational accelerations by $ant(\ddot{\theta})$ and $ag(\ddot{\theta})$, which are a monotonically increasing, resp., decreasing function of $\ddot{\theta}$. These muscle-specific relations also hold for the primary spindle afferent firing rates (PSAFRs) which increase with the intrafusal fiber accelerations.

I now introduce second order neurons that encode a joint's (instead of a muscle's) intrafusal fiber accelerations as follows:

$$FR(\ddot{\theta}) \propto FR\left(ant(\ddot{\theta})\right) - FR\left(ag(\ddot{\theta})\right) \qquad \text{Eq. 4}$$

Thus, these neurons fire in proportion to the difference between the antagonist and the agonist primary afferent firing rates. The existence of such neurons (or functionally equivalent neural structures) is in line with a psychophysics experiment that demonstrated that the perceived joint angle depends on the vibratory stimulation of both the agonist and the antagonist [43]. Eq. 3 can now be expressed in terms of the firing rate of these second order neurons:

$$FR(\ddot{\theta}) \propto c\dot{\theta} + k\theta + M^{-1}z$$

In the above, I have tacitly assumed that the translational acceleration in a muscle-tendon unit is dominated by the acceleration in the muscle's extrafusal fibers. This implies that the tendon may not be so compliant that the muscle's acceleration is absorbed in the tendon. In the Discussion,

I will discuss the consequences of a muscle-tendon unit with a larger tendon compliance.

*How can the CNS deal with reafferent feedback?*

I consider two ways in which the CNS can deal with the reafferent feedback $M^{-1}z$: (1) cancellation at the source (the muscle spindle) by fusimotor control, and (2) correction at the level of the CNS by modelling the reafferent feedback using efference copy information $u$. The cancellation mechanism builds on the following equation for the firing rates of the second order neurons:

$$FR\left(\ddot{\theta} - M^{-1}w\right) \propto c\dot{\theta} + k\theta + noise \qquad \text{Eq. 5}$$

In this equation, the cancellation term $-M^{-1}w$ follows from fusimotor-controlled intrafusal fiber acceleration; under perfectly coordinated fusimotor and skeletomotor control, the net acceleration $\ddot{\theta} - M^{-1}w$ only depends on the non-muscular forces, which is reflected in the fact that there is no reafferent component in the right side of Eq. 5. To allow intrafusal and extrafusal fibers to exhibit a different acceleration noise, I assume $w = u + f \neq u + m = z$, in which $f$ and $m$ are, resp., fusimotor and skeletomotor noise (see Fig. 1A and 1B). As a result, the noise in Eq. 5 equals $M^{-1}(m - f)$.

The correction mechanism builds on the following equation for the firing rates:

$$FR\left(\ddot{\theta}\right) \propto c\dot{\theta} + k\theta + M^{-1}u + noise \qquad \text{Eq. 6}$$

In this equation, $M^{-1}u$ approximates the reafferent acceleration $M^{-1}z$, and the noise equals $M^{-1}m$. Because the CNS has access to an efference copy of $u$, it can model the reafferent acceleration $M^{-1}z$ by $M^{-1}u$, allowing it to extract the exafferent component from the total acceleration feedback $FR\left(\ddot{\theta}\right)$.

The cancellation mechanism follows from perfectly coordinated fusimotor and skeletomotor control, and this may be an unrealistic assumption. Under imperfect coordination, a part of the reafferent acceleration is not cancelled. This motivates the separation of the reafferent acceleration $M^{-1}z$ in a cancelled and a non-cancelled fraction. Denoting the non-cancelled fraction of the reafferent acceleration by $p$ $(0 \leq p \leq 1)$ allows for a hybrid mechanism:

$$FR\big(\ddot{\theta} - (1-p)M^{-1}w\big) \propto c\dot{\theta} + k\theta + pM^{-1}u + noise \qquad \text{Eq. 7}$$

In this hybrid mechanism, the CNS models $pM^{-1}z$, the part of the reafferent acceleration that is not cancelled at the source. The noise is equal to $M^{-1}m - (1-p)M^{-1}f$. In the remainder of this paper, I will mainly focus on the pure cancellation and the pure correction mechanism ($p$ =0 resp. $p$ =1).

*Empirical evidence*

I now present three sets of empirical evidence that support my claims about how the CNS can deal with reafferent feedback. The first set pertains to the fact that reafferent acceleration is not only produced by the spindle-bearing muscle, but by all muscles that act on the same joint, both agonists and antagonists. Therefore, the cancellation mechanism requires that the CNS combines the planned activations of all muscles affecting the same joint and cancels the resulting reafferent feedback via fusimotor neurons. This empirical evidence is reviewed in [7, 8, 37, 38], and the most important findings involve that the PSAFRs are (1) negatively related to antagonist muscle activity [37], (2) adjusted according to visual information about target location [44], and (3) modulated by the changing adaptation state during motor learning [45].

The second set of evidence also pertains to the cancellation mechanism, and the relevant data are the PSAFRs in an active stationary joint. In this scenario, a stationary joint is actively contracting against an external load (isometric contraction). This scenario is typical for standing balance because people prefer a leaned forward posture. In this position, the calf muscles produce a torque that acts against the gravitational torque that results from this posture. In this scenario, $\ddot{\theta} = 0$, $\dot{\theta} = 0$, and $\theta > 0$, and it follows from Eq. 3 (the equation for the cancellation mechanism) that

$$FR(-M^{-1}w) \propto k\theta + noise \qquad \text{Eq. 8}$$

Disregarding the noise, this equation states that the firing rate $FR(-M^{-1}w)$ is proportional to $k\theta$. Because the firing rate depends on the PSAFRs of the joint-crossing muscles (see Eq. 4), this equation implies that the PSAFRs of the muscle that acts against a load also reflects the size of this load. This prediction is confirmed in human microneurography experiments involving stationary joints acting against an external load [30, 31]. These load-dependent PSAFRs are even observed in a slowly contracting muscle [32, 33, 35]. Although the mechanism behind this phenomenon is unknown, it is consistent with the fact that the firing rate reflects the tension in the intrafusal fibers of the muscle that acts against the load.

These results on the PSAFRs in a stationary joint requires a qualification of our earlier statement that PSFARs encode acceleration [10-13]. Crucially, both sets of results (PSFARs in a moving and a stationary joint) are consistent with Eq. 5: PSAFRs are proportional to the exafferent acceleration in a moving joint but also to the exafferent acceleration (i.e., $k\theta$) that would be observed if there was no load-dependent muscular torque that keeps the joint stationary. Thus, I keep my claim that PSFARs encode acceleration but distinguish between an acceleration that is observed and an acceleration that would be observed in the absence of a counteracting muscular torque.

The third set of evidence pertains to the prediction that the PSAFRs depend on both state-dependent and state-independent accelerations. The state-dependent accelerations depend on the state variables $\theta$ and $\dot{\theta}$, and the state-independent accelerations depend on the noise term in Eq. 5. The role of these two types of accelerations has been investigated in studies with unpredictable mechanical perturbations [46-49]. To relate this empirical evidence to the formalisms presented in this paper, I add these unpredictable mechanical perturbations to the noise $m$ in the mechanical system input $z = u + m$. Thus, $m$ does not only contain internally generated noise components, but also unpredictable external mechanical perturbations.

Unpredictable mechanical perturbations have been used to investigate the kinematic signal that drives the corrective response [46-49]. The effect of these perturbations on the spindle afferents develops over time: shortly after perturbation onset, when the state variables $\theta$ and $\dot{\theta}$ are still close to their target values (the stationary upright position), they produce a state-independent intrafusal fiber acceleration, but as time develops, the perturbation also affects the state variables. These late effects on the state variables produce intrafusal fiber accelerations that the CNS can use to estimate the state, allowing it to be controlled. Eq. 5 predicts an early perturbation-induced response that scales with state-independent acceleration, and a later response that scales with the state variables. This is exactly what has been observed at the level of the EMG in a series of experiments on standing balance control under external perturbations, both in cats and in humans [46-49].

*How can the CNS extract state information from the sensory feedback?*

In the cancellation scenario, the sensory feedback is purely exafferent, but the primary spindle afferents do not separate the position and the velocity component into different channels. Instead, they send a mixture of these state variables to the CNS. This is incomplete sensory feedback. To show how the CNS can extract state information from this incomplete feedback, I start from Eq. 5, ignore the noise term, and rewrite it in matrix notation:

$$FR\left(\begin{bmatrix}1 & -M^{-1}\end{bmatrix}\begin{bmatrix}\ddot{\theta}\\ w\end{bmatrix}\right) \propto \begin{bmatrix}k & c\end{bmatrix}\begin{bmatrix}\theta\\ \dot{\theta}\end{bmatrix} = \begin{bmatrix}k & c\end{bmatrix}\boldsymbol{x} \qquad \text{Eq. 10}$$

The coefficient matrix on the left side of Eq. 10 is the green matrix *C* in the sensory system of Fig. 1A and 1B, and the coefficient matrix on the right side is the blue matrix *C* in Fig. 1B.

$$FR\left(C_{Sens}\begin{bmatrix}\ddot{\theta}\\ w\end{bmatrix}\right) \propto C_{Comp}\begin{bmatrix}\theta\\ \dot{\theta}\end{bmatrix} = C_{Comp}\boldsymbol{x} \qquad \text{Eq. 11}$$

I assume that the firing rate function $FR$ is approximately linear over the range of its argument. Under this assumption, Eq. 11 can be written as follows:

$$C_{Sens}\begin{bmatrix}\ddot{\theta}\\ w\end{bmatrix} \approx C_{Comp}\boldsymbol{x} \qquad \text{Eq. 12}$$

in which $C_{Comp}$ absorbs the constant of proportionality. Thus, there exists a linear combination of the sensory feedback variables $\ddot{\theta}$ and $w$ that equals a linear combination of the state variables $\boldsymbol{x}$. The matrix $C_{Comp}$ defines a sensory internal model [50].

In the correction mechanism, the sensory feedback equals the extrafusal fiber acceleration $\ddot{\theta}$ (total acceleration), and the matrix $C_{Sens}$ thus becomes 1 (the identity matrix in the multivariate case; see further). With this new definition of $C_{Sens}$, the following equation can be derived:

$$C_{Sens}\ddot{\theta} \approx C_{Comp}\boldsymbol{x} + Du \qquad \text{Eq. 13}$$

In this equation, I follow the convention of denoting the feedthrough/feedforward component by $Du$; thus, $D = M^{-1}$. Eq. 13 shows how the CNS can use the efference copy $u$ to correct for the reafferent feedback that is a part of $\ddot{\theta}$.

For the hybrid mechanism in Eq. 7 one obtains a mixture of Eq. 12 and Eq. 13: the structure is that of Eq. 13, but $C_{Sens} = [1 \quad -(1-p)M^{-1}]$ (instead of 1 or the identity matrix), $\ddot{\theta}$ is replaced by $\begin{bmatrix} \ddot{\theta} \\ w \end{bmatrix}$, and $D$ is replaced by $pD$.

I now consider the question whether the CNS can estimate the states from a time history of incomplete sensory feedback. This property is known as observability, and it can be assessed from the row space of the so-called observability matrix [51]. Unfortunately, this matrix condition does not tell whether the mechanical system of interest can be controlled with realistic state and control variables (body angular position, angular velocity, and ankle torques) and realistic noise levels. In this paper, I will use simulation to evaluate whether sensory feedback governed by the matrix pair $[C_{Sens}, C_{Comp}]$ or the matrix triplet $[C_{Sens}, C_{Comp}, D]$, for exafferent resp. total acceleration feedback, is sufficient for realistic standing balance control. Crucially, at this point, it is unknown whether total acceleration feedback must be corrected for the reafferent acceleration (using the efference copy $u$ and the matrix $D$) such that enough state information can be extracted. I will evaluate this in simulations with total acceleration feedback and $D = 0$.

## An optimal computational system to evaluate the sufficiency of acceleration feedback

To evaluate whether acceleration feedback is sufficient for standing balance control I follow this rationale: if an optimal computational system with sensory model $C_{Comp}$ or $[C_{Comp}, D]$, for exafferent resp. total acceleration feedback, cannot control the CIP with a realistic body sway and ankle torque, then that acceleration cannot be sufficient for standing balance control. Note that this rationale only holds if the computational system is optimal; if it were suboptimal, then a failure to control standing balance using acceleration feedback could be due to the computational system instead of the feedback type. This type of modelling is fully in line with a recent proposal by [52].

The starting-point are the nonlinear CIP EoM in state-space form, which follows from Eq. 2:

$$\dot{\boldsymbol{x}} = \begin{bmatrix} \dot{\theta} \\ \ddot{\theta} \end{bmatrix} = \begin{bmatrix} \dot{\theta} \\ M^{-1}\left(-c_{damp}\dot{\theta} + \frac{1}{2}mgl\sin(\theta) - k_{stiff}\,\theta + z\right) \end{bmatrix}$$
$$= \Omega(\boldsymbol{x}, \boldsymbol{z})$$

I assume that the CNS tries to control the CIP using an internal forward model that is a linear approximation of $\Omega(\boldsymbol{x}, \boldsymbol{z})$. This linear approximation is obtained by taking the first order Taylor expansion around 0 (for $\theta$, $\dot{\theta}$ and $z$):

$$\dot{\boldsymbol{x}} \approx \begin{bmatrix} 0 & 1 \\ M^{-1}\left(\frac{1}{2}mgl - k_{stiff}\right) & -M^{-1}c_{damp} \end{bmatrix}\boldsymbol{x} + \begin{bmatrix} 0 \\ M^{-1} \end{bmatrix}\boldsymbol{z} \qquad \text{Eq. 14}$$
$$\approx A\boldsymbol{x} + B\boldsymbol{z}$$

Note that $B\boldsymbol{z} = B(\boldsymbol{u} + \boldsymbol{m}) = B\boldsymbol{u} + B\boldsymbol{m}$, which implies that the system noise ($B\boldsymbol{m}$) is a linear function of the noise in its input. This follows from the assumption that the mechanical system is deterministic for a given input $\boldsymbol{z}$.

I will combine the internal forward model in Eq. 14 with two different internal sensory models, one for exafferent and one for total acceleration feedback, corresponding to resp. the cancellation and the correction mechanism. The sensory model for exafferent acceleration feedback is the following:

$$y = C_{Comp}\boldsymbol{x} + noise \qquad \text{Eq. 15}$$

According to this model, the sensory feedback only depends on a single systematic variable, the mechanical system's state $\boldsymbol{x}$. The noise term on the right side of Eq. 15 equals $M^{-1}(\boldsymbol{m} - \boldsymbol{f}) + \boldsymbol{s}$, in which $\boldsymbol{s}$ is pure sensory noise that is added on top of the sensory noise that follows from the motor and the fusimotor noise. This pure sensory noise is added in the process between sensory transduction in the muscle spindle and the arrival of the signal in the CNS.

I now describe linear-quadratic-Gaussian (LQG) control [1] under the dynamical system specified by Eq. 14 and Eq. 15. LQG control is based on the minimization of the following time domain optimality criterion:

$$J = \lim_{T\to\infty} \frac{1}{T} \mathcal{E}\left( \int_0^T \{[\boldsymbol{x}(t) - \boldsymbol{r}]^t Q [\boldsymbol{x}(t) - \boldsymbol{r}] + \boldsymbol{u}(t)^t R \boldsymbol{u}(t)\} dt \right) \qquad \text{Eq. 16}$$

in which $\mathcal{E}(\,)$ denotes expected value, and $Q$ and $R$ are positive definite weight matrices of the appropriate dimensions. This criterion is a quantification of the person's objective to stay balanced with as little effort as possible. The first quadratic form ($[\boldsymbol{x}(t) - \boldsymbol{r}]^t Q [\boldsymbol{x}(t) - \boldsymbol{r}]$) quantifies the difference between the time-varying state variable $\boldsymbol{x}(t)$ and its target state $\boldsymbol{r} = [a^+, 0]^t$ (in which $a^+$ is a small positive number that reflects people's preference for a leaned forward posture), and the second one ($\boldsymbol{u}(t)^t R \boldsymbol{u}(t)$) quantifies the energetic cost.

The minimization of $J$ has a known solution for (1) a linear dynamical system (e.g., the one specified by Eq. 14 and Eq. 15), and (2) Gaussian sensor and motor noise. Optimal control under the LQG criterion is governed by a state observer, a linear differential equation that describes the dynamics of the state estimates $\hat{\boldsymbol{x}}$:

$$\dot{\hat{\boldsymbol{x}}} = A\hat{\boldsymbol{x}} + B\boldsymbol{u} + L(\boldsymbol{y} - C_{Comp}\hat{\boldsymbol{x}}) \quad \text{Eq. 17}$$

in which $L$ is the Kalman gain [1]. Optimality also requires that the control action $\boldsymbol{u}$ is obtained as $\boldsymbol{u} = -K(\hat{\boldsymbol{x}} - \boldsymbol{r})$, in which $-K$ is the linear quadratic regulator (LQR) gain [1]. The LQR gain depends on $A$ and $B$ (plus the weights $Q$ and $R$), and the Kalman gain depends on $A$ and $C$ (plus three noise covariance matrices, which serve as weights) [1]. The variable $\boldsymbol{u}$ in Eq. 17 is an efference copy of the control signal that is sent to the muscles.

Eq. 17 is the standard solution of the LQG with a few adaptations to the usual way of calculating the Kalman gain: (1) the system noise covariance matrix depends on $B$, (2) the sensor noise covariance matrix not only depends on pure sensory noise but also on the motor and the fusimotor noise (via the term $M^{-1}(\boldsymbol{m} - \boldsymbol{f})$), and (3) the system and the sensor noise are correlated (because the motor noise $\boldsymbol{m}$ appears in both the state and the output equation and is assumed to be correlated with the fusimotor noise).

In the absence of fusimotor control, the CNS receives total acceleration feedback, which corresponds to the following output equation:

$$y = C_{Comp}\boldsymbol{x} + D\boldsymbol{u} + noise$$

Crucially, this output equation contains a systematic term ($D\boldsymbol{u}$) that adds reafferent acceleration to the exafferent acceleration $C_{Comp}\boldsymbol{x}$. The noise term is equal to $M^{-1}\boldsymbol{m} + \boldsymbol{s}$. The corresponding state observer is the following:

$$\dot{\hat{\boldsymbol{x}}} = A\hat{\boldsymbol{x}} + B\boldsymbol{u} + L(\boldsymbol{y} - C_{Comp}\hat{\boldsymbol{x}} - D\boldsymbol{u}) \quad \text{Eq. 18}$$

This is a state observer with correction for reafference, which is performed by $D\boldsymbol{u}$. To prevent clutter, this part of the state observer is not shown in Fig. 1.

For completeness, the sensory model for the hybrid mechanism contains the term $p\boldsymbol{Du}$ instead of $\boldsymbol{Du}$, and its noise term is $M^{-1}\boldsymbol{m} - (1-p)M^{-1}\boldsymbol{f} + \boldsymbol{s}$. Because the Kalman gain depends on a sensor noise covariance matrix, for the hybrid mechanism it depends on the parameter $p$.

It is useful to evaluate the contribution of the reafference correction term $-\boldsymbol{Du}$ in the state observer in Eq. 18. With total acceleration feedback, $\boldsymbol{Du}$ is always a part of the feedback but, in principle, the computational system may or not include the corresponding correction term $-\boldsymbol{Du}$ in the state observer. After dropping the correction term from the state observer and some simple algebra, it can be shown that the estimation error $\boldsymbol{x} - \hat{\boldsymbol{x}}$ is governed by the following differential equation:

$$\dot{\boldsymbol{x}} - \dot{\hat{\boldsymbol{x}}} = \left(A - LC_{Comp}\right)(\boldsymbol{x} - \hat{\boldsymbol{x}}) + L\boldsymbol{Du} + noise$$

If the correction term $-\boldsymbol{Du}$ is a part of the state observer, there is no term $L\boldsymbol{Du}$ in this estimation error differential equation. Crucially, the Kalman gain $L$ is such that all eigenvalues of the matrix $\left(A - LC_{Comp}\right)$ have negative real parts. This guarantees that, with the correction term $-\boldsymbol{Du}$ in the state observer, $\hat{\boldsymbol{x}}$ converges to $\boldsymbol{x}$. This guaranteed convergence is an important advantage of reafference correction, but it does not imply that balance control is impossible or substantially worse without reafference correction. To address this issue, I will use simulations in which I compare the control performance of an observer with and without reafference correction.

### Standing balance can be controlled using ankle joint acceleration feedback

I now report on a simulation study that evaluates whether acceleration feedback is sufficient to control standing balance with realistic model parameters and realistic kinematic simulation output. In this section, I give a general description of the simulation study; details are given in the Methods. First, the mechanical system (the CIP) is specified by realistic and/or empirical values for its parameters (body length and mass, stiffness, damping). Second, the simulations require noise input, and I set the amplitude of the motor noise (skeletomotor plus fusimotor, if present) such that its effect on the sensory feedback matches the effect of the pure sensory noise. If fusimotor noise is present (only when simulating exafferent acceleration feedback), then its amplitude is equal to the amplitude of the skeletomotor noise. The shared variance of the motor and the fusimotor noise is 25%. Third, the LQR gain depends on the weights of the optimality criterium in Eq. 16 (the 2-by-2 matrix $Q$ and the scalar $R$) and these are set such that the precision ($Q$-dependent) and the energetic cost ($R$-dependent) component have an equal contribution to the optimality criterium.

In the human neuromuscular system, every joint has a torque limit, and I therefore ran simulations in which the output of the computational system was truncated at 195 Nm, which is an empirically determined maximum ankle torque [53].

I evaluate the model's performance using the angular position root-mean-square deviation (RMSD). For a balance control model to be plausible, (1) the angular position RMSD must converge to zero with decreasing noise amplitude, and (2) the whole RMSD range that is observed in humans must be produced by the model. The maximum RMSD was obtained from a study in which participants had to rely on proprioceptive feedback only (vestibular loss patients with their eyes closed) [54]; these participants did not tolerate an angular position RMSD above 1.5 degrees [54, Fig. 4].

Crucially, this maximum RMSD must be produced while the mechanical system's CoG stays within its AoS (see Methods).

Simulation results are shown in Fig. 2, separately for exafferent feedback (panels A and C) and total acceleration feedback with correction for reafference (panels B and D). Panels A and B show that (1) the angular position RMSD converges to zero with decreasing noise amplitude, and (2) the maximum tolerable RMSD is reached for a smaller noise amplitude than the one for which the CoG moves outside the AoS. To interpret the results in Fig. 2, it is not necessary to know the noise amplitudes, because the only relevant variables are the kinematic and the motor output variables, which must remain within a plausible range. In fact, there is no empirical reference value for these noise amplitudes. The results in Fig. 2 were obtained using motor and sensory noise amplitudes that had an equal impact on the sensory feedback (see Methods - Simulations of standing balance control). However, these results are robust to variations in the relative noise amplitudes (motor versus sensory) within an order of magnitude. The results are also robust to variations in the relative weights of the LQR optimality criterium (precision versus energetic cost), and this over more than an order of magnitude.

Simulations were also run with the hybrid mechanism, and qualitatively similar results were obtained: for all values of the parameter $p$ (see Eq. 7), the angular position RMSD converges to zero with decreasing noise amplitude, and the maximum tolerable RMSD is reached for a smaller noise amplitude than the one for which the CoG moves outside the AoS.

Panels A and B in Fig. 2 give no information about the frequency content of the angular position time series. However, a rough idea about this frequency content can be obtained from the example time series in Panels

C and D. Example time series for other noise levels look highly similar but have amplitudes that scale with the noise level.

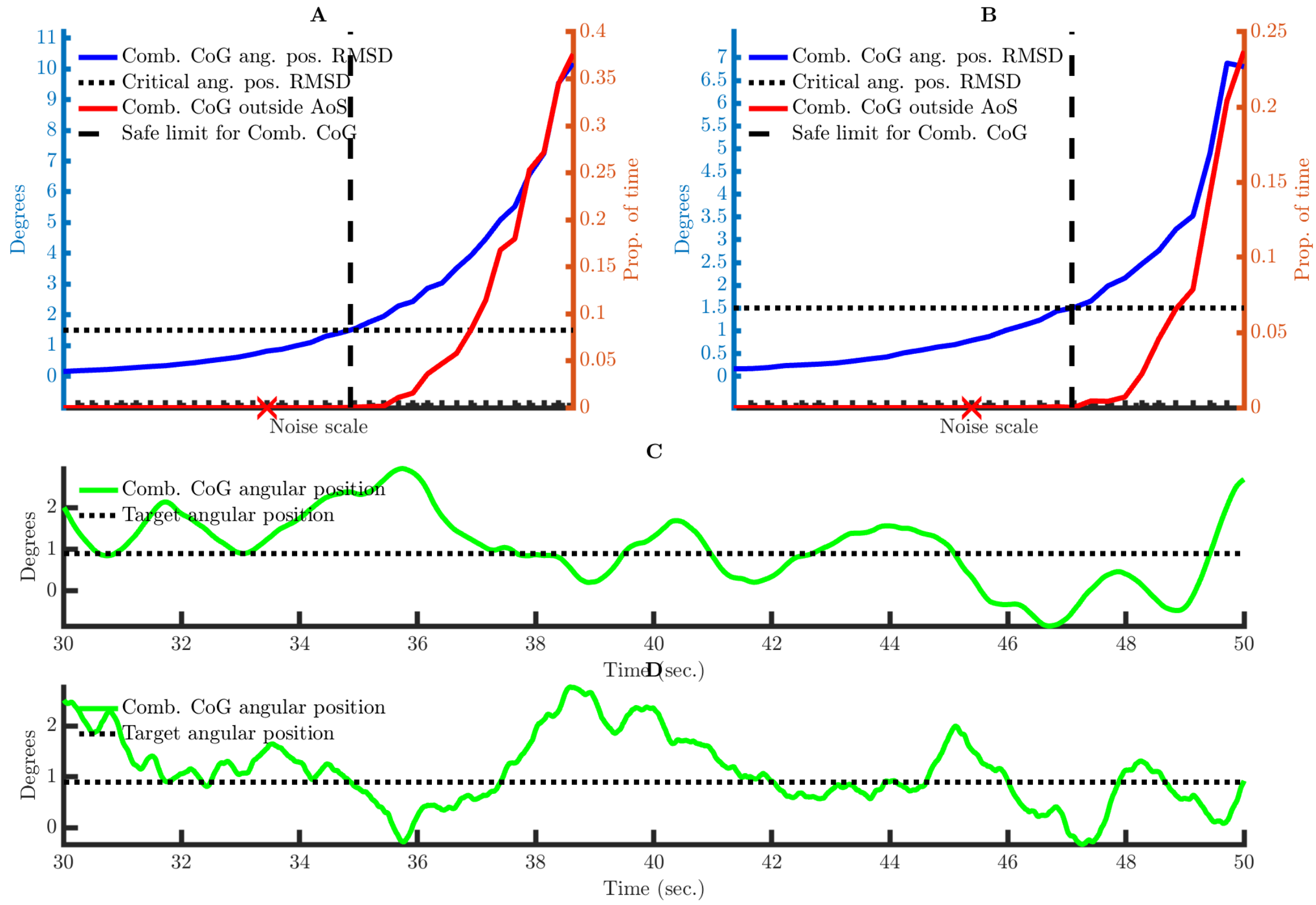

**Fig. 2: Simulation results for standing balance control using acceleration sensory feedback.** (A&B) Two indices of lack of control as a function of noise: angular position RMSD and the proportion of the time that the angular position exceeds its critical value. The horizontal dotted line indicates the maximum RMSD based on [54], and the vertical dashed line indicates the first noise amplitude for which the computational system cannot keep the mechanical system's CoG within its AoS. (C&D) Example angular position time series for the noise scales indicated by a red cross in panel A and B. Panel A and C show the results for exafferent acceleration feedback, and panel B and D for total acceleration feedback with correction for the reafference.

Totally different results were obtained for total acceleration feedback without correction for reafference: for all noise levels, the model produced diverging lean angles. The lowest noise level tested was close to the minimum noise level that is required to compute the Kalman gain. Although this is not a mathematical proof of the claim that reafferent feedback must be corrected to control balance successfully, the evidence is as good as one can get using numerical methods.

This simulation study shows that joint acceleration feedback is sufficient for standing balance control (i.e., the computational system keeps the CoG angular position in a realistic regime), but only if the reafferent acceleration component is either absent from the feedback (cancellation mechanism) or it is corrected for in the state observer (correction mechanism).

## Bicycle balance control

### *Problem definition and bicycle mechanics*

There are two important differences between standing and bicycle balance control: (1) a bicycle's AoS is a line instead of a surface, and (2) balance control of a moving bicycle involves not only the gravitational but also the centrifugal force. A stationary bicycle is balanced when the combined CoG of rider and bicycle is exactly above the line that connects the contact points of the two wheels with the road surface, the so-called line of support (LoS). On a moving bicycle, not only gravity, but also the centrifugal force acts on the CoG. Crucially, the centrifugal force is under the rider's control via the turn radius [55]. The balance of a moving bicycle depends on the resultant of all forces that act on the CoG: a moving bicycle is balanced if the resultant of the gravitational and the centrifugal force is directed towards the LoS. Because of disturbances, the direction of this resultant force cannot be exactly towards this one-dimensional LoS for some time. Therefore, a bicycle is considered balanced if this resultant force fluctuates around the LoS within a limited range, small enough to prevent the bicycle from touching the road surface.

Besides the forces that act on the CoG, there are also forces that turn the bicycle's front frame, and some of these forces are independent of the rider's actions [56]. These rider-independent forces are responsible for the bicycle's self-stability and will not be described in detail.

For investigating bicycle balance control, I use the bicycle model in Fig. 3. This model consists of three rigid bodies: front frame, rear frame, and the rider's upper body. The rear frame includes the lower body and will therefore also be denoted as the lower body. The positions of these three bodies are specified by three angular variables: steering ($\delta$), lower body ($\theta_1$), and upper body ($\theta_2$) angular position. The lower and upper body angular positions are relative to gravity.

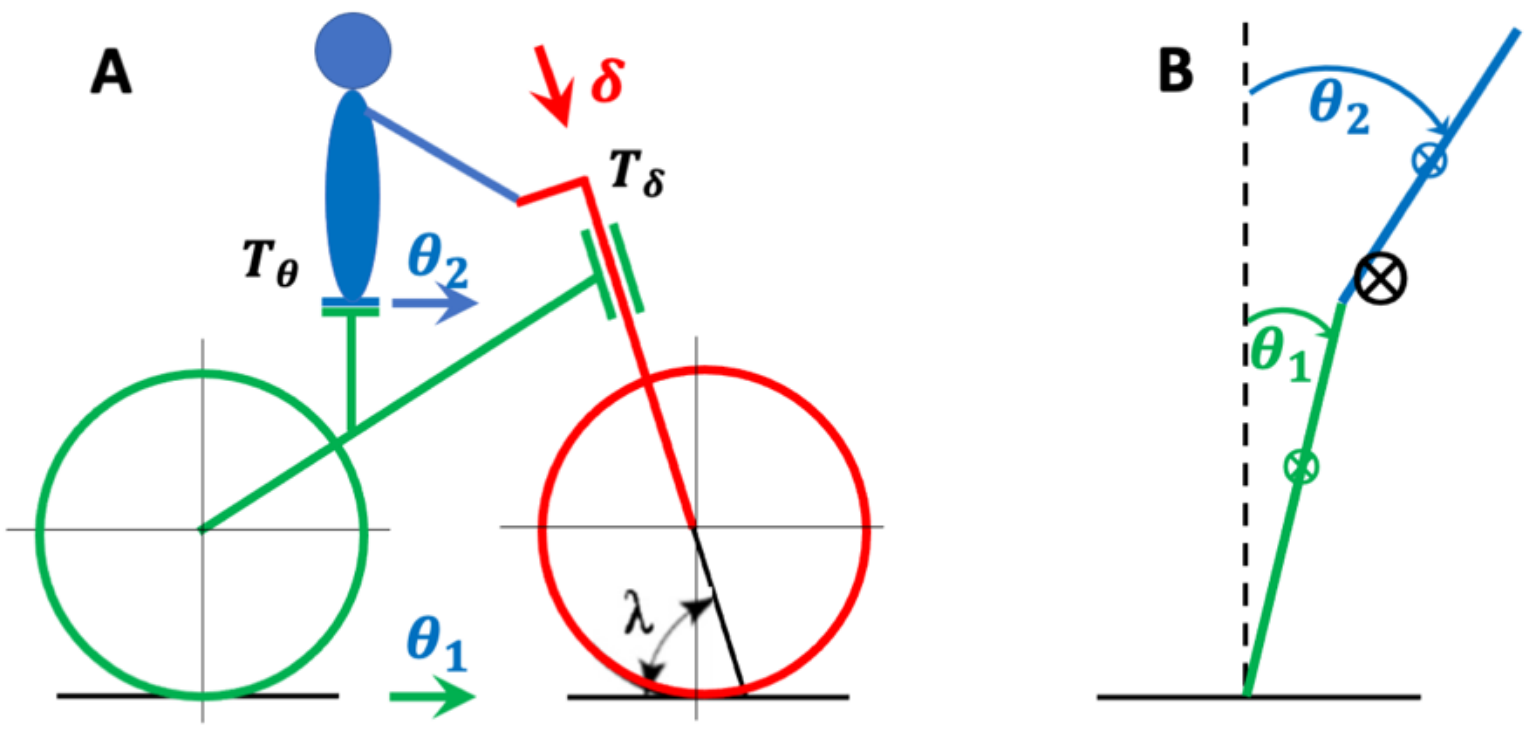

**Fig 3: Kinematic variables of the bicycle model plus the rider-controlled forcing torques.** (A) Side view. In green, the bicycle rear frame, characterized by its angular position $\theta_1$ over the roll axis (green arrow). In red, the bicycle front frame, characterized by its angular position $\delta$ over the steering axis (red arrow). In blue, the rider's upper body, characterized by its angular position $\theta_2$ over the roll axis (blue arrow). In black, (1) the steering torque $T_\delta$ and the pelvic torque $T_\theta$, which are both applied by the rider, and (2) the steering axis angle $\lambda$ (see text). (B) Rear view. In green, the bicycle rear frame (including the lower body) angular position $\theta_1$. In blue, the rider's upper body angular position $\theta_2$. The symbol $\otimes$ denotes the CoG of the upper body (in blue), the lower body (in green), and the combined CoG (in black).

Cycling involves a double balance problem, of which I have only described the first part, which is keeping the combined CoG of rider and bicycle above the LoS. The second balance problem pertains to the rider's upper body only, and it involves keeping the upper body CoG above its AoS, the saddle. I will only consider the balance over the roll axis (parallel to the LoS), which corresponds to upper body movements to the left and the right. I will thus ignore the balance over the pitch axis (perpendicular to the LoS and gravity), which corresponds to upper body movements to the front and the back, typically caused by accelerations and braking.

For both balance problems (with respect to the combined and the upper body CoG), the relevant control actions must result in a torque over the roll axis. Within the constraints of our kinematic model, there are two possible control actions: (1) turning the handlebars (using steering torque $T_\delta$), and (2) leaning the upper body (using pelvic torque $T_\theta$). At this point, it is convenient to make use of Fig. 3B, which is a schematic of a double compound pendulum (DCP). By turning the handlebars, the contact point of the front tire (represented by the green rod) with the road surface moves to the left or the right, and this changes the position of the combined CoG relative to the LoS. In the bicycle reference frame (in which the LoS is one of the axes) this corresponds to a centrifugal torque in the direction opposite to the turn (a tipping out torque). Steering in the direction of the lean produces a tipping out torque that brings the combined CoG over the LoS. This is called steering to the lean/fall.

The second control action is leaning the upper body, which can bring the upper body CoG above the saddle in a direct way. This deals with the second balance problem. However, there is consensus that leaning the upper body cannot deal with the first balance problem (bringing the combined CoG above the LoS), at least not in a direct way. An important argument in favor of this view is that a bicycle with a locked steer cannot be balanced; not a single case has been reported. However, leaning the upper body can deal with the first balance problem in an indirect way: leaning the upper body to one side will make the front and the rear frame lean to the other side (by conservation of angular momentum). Depending on geometrical properties of the bicycle, this front and rear frame lean (i.e., roll angle) may turn the front frame (i.e., change its yaw angle) to the same side [56, 57].

*What sensory feedback informs the CNS about the combined CoG?*

One of the most challenging aspects of bicycle balance control pertains to the sensory feedback that informs the CNS about the combined CoG angular position and velocity, which is a function of the states of the combined rider-bicycle mechanical system. To describe this, I use a relation that holds for a general multibody mechanical system, and that is derived in the Methods:

$$[I, -M(\mathbf{0})^{-1}H]\begin{bmatrix}\ddot{\boldsymbol{\theta}}\\ \boldsymbol{z}\end{bmatrix} \approx J_{[\boldsymbol{\theta};\dot{\boldsymbol{\theta}}]}(\mathbf{0};\mathbf{0})\begin{bmatrix}\boldsymbol{\theta}\\ \dot{\boldsymbol{\theta}}\end{bmatrix}$$

The symbols in this equation map onto the ones in Fig. 5 as follows: $\boldsymbol{\theta} = [\delta, \theta_1, \theta_2]^t$ and $\boldsymbol{z} = [T_\delta, T_\theta]^t$. In this equation, $M(\mathbf{0})$ is a mass moment of inertia, $H$ is a matrix that distributes the forcing torques over the joints, and $J_{[\boldsymbol{\theta};\dot{\boldsymbol{\theta}}]}(\mathbf{0};\mathbf{0})$ is the Jacobian of the EoM. In the same way as for the CIP, I assume noisy intrafusal fiber accelerations $M(\mathbf{0})^{-1}H\boldsymbol{w}$ that approximate the reafferent acceleration feedback $M(\mathbf{0})^{-1}H\boldsymbol{z}$. Thus, under perfectly coordinated (but noisy) fusimotor and skeletomotor control, one obtains the following model for exafferent acceleration feedback:

$$[I, -M(\mathbf{0})^{-1}H]\begin{bmatrix}\ddot{\boldsymbol{\theta}}\\ \boldsymbol{w}\end{bmatrix} \approx J_{[\boldsymbol{\theta};\dot{\boldsymbol{\theta}}]}(\mathbf{0};\mathbf{0})\begin{bmatrix}\boldsymbol{\theta}\\ \dot{\boldsymbol{\theta}}\end{bmatrix}$$

$$C_{Sens}\begin{bmatrix}\ddot{\boldsymbol{\theta}}\\ \boldsymbol{w}\end{bmatrix} \approx C_{Comp}\begin{bmatrix}\boldsymbol{\theta}\\ \dot{\boldsymbol{\theta}}\end{bmatrix}$$

Without fusimotor control, the primary spindle afferents produce total acceleration feedback, which is modeled as follows:

$$I\ddot{\boldsymbol{\theta}} \approx J_{[\boldsymbol{\theta};\dot{\boldsymbol{\theta}}]}(\mathbf{0};\mathbf{0})\begin{bmatrix}\boldsymbol{\theta}\\ \dot{\boldsymbol{\theta}}\end{bmatrix} + M(\mathbf{0})^{-1}H\boldsymbol{z}$$

$$C_{Sens}\ddot{\boldsymbol{\theta}} \approx C_{Comp}\begin{bmatrix}\boldsymbol{\theta}\\ \dot{\boldsymbol{\theta}}\end{bmatrix} + D\boldsymbol{u}$$

Thus, $C_{Sens}$ is an identity matrix, and the feedforward term $D\boldsymbol{u}$ models the reafferent acceleration component.

It is important to note that the joint between the lower body (rear frame) and the road surface is not actuated, and this is captured by the matrix $H$. Thus, there is no torque that corresponds to $\theta_1$, and the torque $T_\theta$ is the pelvic torque that corresponds to $\theta_2$. Because our bicycle model has less actuators than degrees of freedom, it belongs to the category of underactuated systems [16]. To control a non-actuated joint, an underactuated system must rely on the actuated joints, and this may be a challenging task for the controller.

As an aside, it must noted that the lower body is also used for propulsion of the bicycle. However, it cannot be ruled out that resting on the pedals allows to register balance-relevant sensory information. For the purpose of this paper, I consider a restricted form of cycling in which the rider keeps his legs still, does not use them to carry weight, and relies on a motor for propulsion. This is like balance control on a scooter.

The matrices $C_{Sens}$ and $C_{Comp}$ all have three rows. Via the term $\boldsymbol{I\ddot{\theta}}$, every row in $C_{Sens}$ and $C_{Comp}$ corresponds to one of the three joints of the bicycle model: the steering joint ($\ddot{\delta}$), the joint between the road and the lower body ($\ddot{\theta}_1$), and the pelvic joint ($\ddot{\theta}_2$). With respect to proprioceptive feedback at the pelvic joint, only the difference acceleration $\ddot{\theta}_2 - \ddot{\theta}_1$ can be registered (see Fig. 3B). This fact requires matrices $C_{Sens}$ and $C_{Comp}$ with two rows, of which one corresponds to $\ddot{\delta}$ and the other to the difference acceleration $\ddot{\theta}_2 - \ddot{\theta}_1$. This can be realized by pre-multiplying the original $C_{Sens}$ and $C_{Comp}$ by the matrix

$$\begin{bmatrix} 1 & 0 & 0 \\ 0 & -1 & 1 \end{bmatrix}$$

Thus, no unique information about the angle between the road and the lower body enters the CNS; the corresponding joint is not only non-actuated, but it also sends no unique information to the CNS. As will be argued in the next paragraph, in the simulations, I will only use the second row of the matrix above (i.e., the vector $[0 \quad -1 \quad 1]$).

I will test the hypothesis that a bicycle can be balanced using the same type of proprioceptive feedback as for standing balance control. The latter involves ankle joint accelerations over the same axis as the angular position that the person wants to control. In bicycle balance control, there are two controllable joints, the steering axis and the pelvis, but only the latter is over the same axis as the CoG angular position that the rider wants to control. Thus, to test the hypothesis that a bicycle and a standing body can be balanced using the same type of feedback, I ignore the acceleration feedback from the steering axis. In my simulations, this was implemented by pre-multiplying $C_{Sens}$ and $C_{Comp}$ by the vector $[0 \quad -1 \quad 1]$. As expected, including the steering axis feedback in the simulations improved the performance of the controller (results not shown). The reported simulation results are thus for a suboptimal scenario, but with the same type of feedback as for standing balance control.

To the best of my knowledge, there is no empirical evidence for the hypothesis that pelvic joint proprioceptive feedback is sufficient for bicycle balance control. However, there are observations that suggest that a bicycle cannot be balanced if the pelvic joint proprioceptive feedback cannot be processed correctly using the rider's internal models. These observations come from a study in which riders were instructed to take turns with a special bicycle (a bricycle) that, on initiation of a turn, tips the rear frame to the outside of the turn [55, 58]. On a bricycle, this happens because, due to a special spring construction, the gravitational torque is cancelled. On a regular bicycle, a turn-induced tipping-out centrifugal torque is cancelled by a lean-induced tipping-in gravitational torque, but on a bricycle no gravitational torque is available to cancel the centrifugal torque [55, 58]. Crucially, none of the participants was able to complete a simple obstacle course on the bricycle [55, 58]. This suggests that the non-negotiable tipping-out torque elicits proprioceptive feedback that makes the CNS think the rider is falling to the opposite side of the initiated turn. Because the CNS has learned to steer to the fall, completing the turn is not possible on the bricycle.

*Two bicycle models*

Starting from the kinematic bicycle model in Fig. 3, I formulated two sets of EoM for the mechanical system, one nonlinear and one linear. The EoM for both bicycle models are derived in the Methods. The nonlinear EoM are obtained by combining the dynamics of the Acrobot and the double pendulum on a cart under bicycle-specific kinematic constraints [59]. This model is called the Steered Double Pendulum (SDP). The linear EoM are obtained from a linear 2-DoF benchmark model [56] by replacing the rear frame by a linearized double pendulum [59]. This linear model is called the Benchmark Double Pendulum (BDP).

Simulations using the linear BDP have the advantage that this model captures the rider-independent steering torques that are present in commercial road bicycles. These involve passive steering torques that depend on the rear frame lean angle, which are necessary for the self-stability of the bicycle [56]. In contrast, the SDP has no passive steering torques and therefore has a much twitchier steering behavior (no resistance to turning). This twitchy steering behavior is due to the SDP's vertical steering axis. Bicycles for artistic cycling have a steep (but not vertical) steering axis, and they are always ridden in a near-vertical position (no lean, wide turns). In the BDP but not in the SDP, the rider can in principle feel the angular position through the steering torque, but this potential feedback has been removed from $C_{Sens}$ and $C_{Comp}$ by pre-multiplying these matrices with $[0 \quad -1 \quad 1]$; the 0 in the first position corresponds to the steering axis.

Simulations using the nonlinear EoM of the SDP have the advantage that they mimic the fact that an internal model (linear in our simulations) typically is only an approximation of a nonlinear mechanical system. Simulations using linear EoM will not capture the inevitable differences in the dynamics of the internal and the mechanical model, and a good control performance may thus give an overly optimistic picture.

*A bicycle can be balanced using pelvic joint acceleration feedback*

I followed the same approach as in the simulations of standing balance control, but now the challenge is much more difficult: I try to control a 3-DoF underactuated (instead of a 1-DoF fully actuated) mechanical model using acceleration feedback from a single joint. As for standing balance control, I set the amplitude of the motor noise (skeletomotor plus fusimotor, if present) such that its effect on the sensory feedback matches the effect of the pure sensory noise. The only difference is that the motor noise is two-dimensional, with a steering and a pelvic component. I assume that the two noise components are independent and have equal variance.

The weights of the optimality criterium in Eq. 16 are calculated in the same way as for standing balance control but now using maxima for the lean and the steering angle, the pelvic and the steering torque. Compared to standing balance control, torque limits are much less important in bicycle balance control. This is because (1) turning the handlebars requires very little force, and (2) rotation of the upper body depends on strong axial muscles that can produce a large pelvic torque. As a maximum, I took the torque that is required to keep the upper body in a horizontal position on a Roman chair (a bed with an unsuspended upper body), but this maximum was never reached in the simulations.

Different from standing balance control, which involves a finite length AoS, there is no realistic critical lean angle for which the combined CoG can no longer be supported by steering the LoS underneath it. In fact, for a realistic speed (15.5 km/h, the speed used in the simulations), and with a sufficient resistance between the tires and the road surface (no skidding), the maximum lean angle that can be corrected by steering is 62 degrees, which is much more than humans tolerate.

Also different from standing balance control, there is no experimentally determined critical CoG lean angle beyond which cyclists feel too uncomfortable, and I therefore used a rougher measure. I calculated the lean angle that was required to take a 7 m. diameter corner at a speed of 15.5 km/h. and found this to be 15 degrees inwards to the turn. Based on informal experiments with experienced amateur mountain bikers, this cornering task is feasible but challenging. I expect that for less experienced cyclists, this lean angle will feel uncomfortable. This critical lean angle is specific for the BDP, which has a similar geometry as a mountain bike. Therefore, for the BDP, I calculated the noise amplitude for which this critical lean angle was exceeded in more than 1 percent of the time.

In the same way as for standing balance control, I evaluated the performance of the model for bicycle balance control based on the following criteria: (1) the CoG lean angle RMSD must converge to zero with decreasing noise amplitude, and (2) the whole RMSD range that can be observed must also be produced by the model. In Fig. 4, the results are shown for the BDP as controlled using exafferent feedback and total acceleration feedback with correction for reafference. For both control scenario's, the CoG lean angle RMSD converges to zero with decreasing noise amplitude, and it covers a range that well exceeds the lean angles at which only experienced cyclists are comfortable.

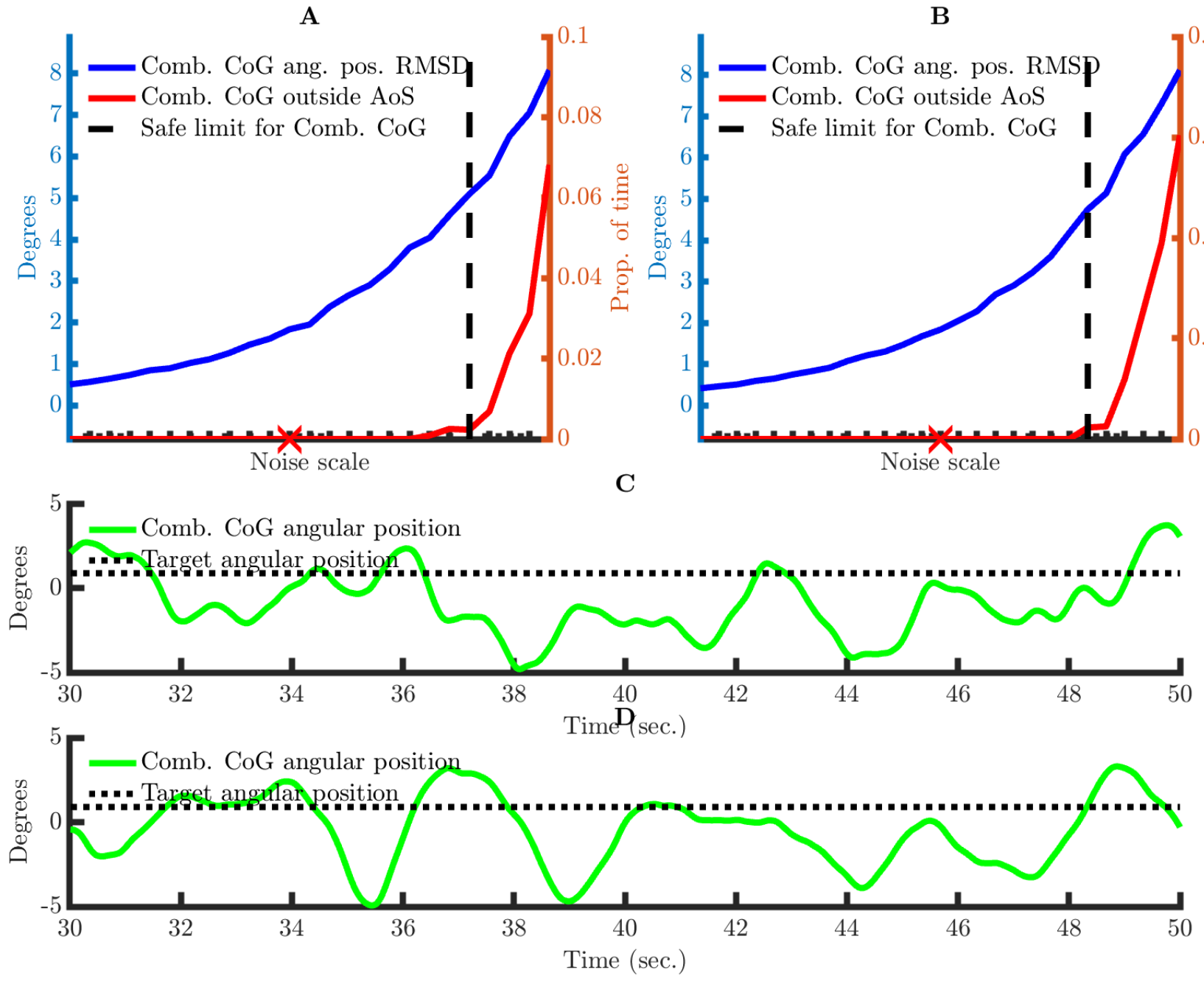

**Fig. 4: Simulation results for BDP bicycle balance control using acceleration sensory feedback.** (A&B) Combined CoG angular position RMSD and the proportion of the time that the angular position exceeds its critical value. (C&D) Example angular position time series for the noise scales indicated by a red cross in panel A and B. Panel A and C show the results for exafferent acceleration feedback, and panel B and D for total acceleration feedback with correction for reafference.

As for standing balance control, without correction for reafference, total acceleration feedback produced diverging lean angles for all noise levels for which a Kalman gain could be computed. This is the strongest numerical evidence possible for the claim that, if reafferent feedback enters the CNS, it must be corrected for.

Different results were obtained for the nonlinear SDP with the vertical steering axis. First, the SDP could only be balanced using exafferent acceleration feedback; total acceleration feedback, with or without correction for reafference, produced diverging lean angles for all noise levels for which a Kalman gain could be computed. Second, for exafferent acceleration feedback, although the CoG lean angle RMSD converges to zero with decreasing noise amplitude, it covers a much smaller range of lean angles than what is possible on a bicycle with BDP geometry: the BPD lean angles diverge when their RMSD exceeds 8 degrees, whereas for the SPD this happens already when the RMSD exceeds 4.2 degrees,

Simulations were also run with the hybrid mechanism, and qualitatively similar results were obtained for both the BDP and the SDP bicycle model. For the SDP, at every noise scale, with increasing parameter $p$, there was a value for $p$ at which the lean angles diverged.

In sum, as for standing balance control, the simulations show that joint acceleration feedback is sufficient for bicycle balance control, but only if the reafferent acceleration component is either absent from the feedback or is corrected for. In addition, there is an advantage of removing the reafferent acceleration feedback already at the level of the muscle spindle: the SDP could be balanced using exafferent but not using total acceleration feedback with correction for reafference.

# Discussion

### Summary of the results

Although there is much debate in sensory neurophysiology about whether muscle spindles provide full state feedback, this issue does not have to be crucial for balance control. This is because a mechanical system can also be controlled using incomplete feedback, and for a linear mechanical system it is known how to do this optimally. This optimal control depends on a CNS-internal dynamical system that estimates the states from the incomplete sensory feedback. This computational system can be implemented as a RNN and it uses a sensory model to update the state estimates using the incomplete feedback. This formalism is highly relevant for spindle primary afferents, whose firing rates scale with acceleration. If fusimotor and skeletomotor control are perfectly coordinated, reafferent acceleration feedback is cancelled and the firing rates of second order (joint level) neurons scale with the exafferent joint acceleration that is produced by the non-muscular forces gravity, elasticity, and damping. In the absence of fusimotor control, these firing rates scale with the sum of the exafferent and the reafferent joint acceleration. Crucially, there exists a sensory model for both components of the feedback: (1) the exafferent acceleration can be expressed as a linear combination of the state variables, and (2) the reafferent acceleration can be expressed as a linear combination of the efference copies of the torque output, and this allows to correct for it. Simulations of standing and bicycle balance control show that joint acceleration feedback is sufficient for balance control, but only if the reafferent acceleration component is either absent from the feedback or is corrected for. This agrees with the fact that the estimation error differential equation converges to 0 only if the reafferent acceleration component is cancelled or corrected for in the state observer. Finally, for a challenging mechanical system like a bicycle with a vertical steering axis, there is an advantage of cancelling the reafferent acceleration feedback already at the level of the muscle spindle.

### Contribution to the modelling of balance control

Several papers in the balance control literature have used an internal dynamical system to model the CNS within the framework of OFC [19, 60-64]. It has been used for testing hypotheses about (1) the information in the sensory feedback (position, velocity, and/or acceleration) that is used for updating the state estimate [60, 61], (2) a mechanism that compensates for the sensorimotor delay [19], (3) intermittent versus continuous standing balance control [62], (4) non-linear components in the corrective responses to balance-perturbing stimuli [63], and (5) the control objective of the CNS (stabilization versus sway minimization) [64]. The main difference with this paper is that I use this framework to test a hypothesis that is derived from sensory neurophysiology: Is joint acceleration feedback sufficient for balance control? Whereas existing papers mainly demonstrate the scope of OFC in explaining different phenomena [19, 60-64], this paper demonstrates that it continues to be successful if the sensory feedback is constrained by the response properties of the relevant sensory organs.

### Short-range muscle stiffness and the mechanics of muscle spindles

Because mechanical systems are second order (i.e., the highest derivative in the EoM is 2), acceleration has a well-defined relation with the state variables. However, there is no guarantee that the mechanics of the muscle spindles have the same structure as the mechanics of the limb to which these spindles belong. In fact, there is good evidence for short-range muscle stiffness, and this potentially complicates the relation between the state variables and the PSAFRs [11, 65-67]. To deal with short-range muscle stiffness, [11] proposed that spindle output scales with a linear combination of the tension force and its first time derivative (yank). However, this same study also demonstrated that, in passive muscles, peak yank and peak acceleration predicted similar amounts of variance in the initial burst amplitudes [11]. Short-range muscle stiffness is thus consistent with spindle output at stretch onset that scales with peak acceleration. This was also demonstrated by a recent combined biomechanical-neurophysiological model of the muscle spindle that models short-range muscle stiffness at the level of cross-bridge kinetics [68].

**Paradoxical muscle shortening**

Joint acceleration feedback combined with sensory-model-based state estimation effectively deals with the potential problems caused by paradoxical muscle shortening [69-71]. Paradoxical muscle shortening is due to compliance of the Achilles tendon, which makes the body unstable, and necessitates muscle activity to maintain balance. Generating sufficient muscular tension results in paradoxical changes in muscle length: when the body rotates forward ($\dot{\theta} > 0$), the calf muscles are actively shortened (producing a negative torque $\tau < 0$) to maintain balance. As a result of this paradoxical muscle shortening, muscle length is negatively correlated with angular position [72]. If PSAFRs would be proportional to muscle length, paradoxical muscle shortening would result in firing rate patterns that are opposite to those of a passive joint that is moved by an external torque. It is unclear how the CNS could interpret such a signal, and this was also noted by [72]. This problem does not exist if the CNS can rely on a sensory internal model to update a state estimate using joint acceleration feedback, as demonstrated in this paper.

At this point, it is important to discuss the sensory consequences of tendon compliance: the larger the compliance, the less the joint angular acceleration will reflect the extrafusal fiber acceleration. This would reduce the information in primary afferent output (which depends on the extrafusal fiber acceleration) about the joint's state. A principled way to deal with a possible inaccurate information about the joint's state is by means of a dynamical system for the muscle-tendon unit in which the muscle and the tendons have their own dynamics. Dynamical systems have been used before to model the properties of sensory organs [18, 19, 62, 63, 73]. Such a dynamical system would become part of an extended mechanical system with additional output variables: lengths, velocities, and accelerations of the individual muscles (extrafusal fibers) and their corresponding tendons. The extrafusal fiber accelerations then co-determine (together with the fusimotor output to the intrafusal fibers) the muscle spindle firing rates. Note that, if the mechanical system is extended by a component for the muscle-tendon dynamics, also the computational system is extended by an internal model for this component.

## Different mechanisms for preventing reafferent feedback from affecting state estimation

I have demonstrated that, for successful balance control, the CNS must prevent reafferent feedback from affecting state estimation. I have described two mechanisms that can achieve this: cancellation by means of temporally coordinated fusimotor and skeletomotor control, and correction using efference copy information. Both mechanisms require learning, but the correction mechanism may be easier. In fact, there is no difference between learning the state observer with or without the correction term $-\boldsymbol{LDu}$: without this correction term, the CNS must learn $\boldsymbol{Bu}$, and with this correction term it must learn $(B - LD)\boldsymbol{u}$ (see Eq. 18). There is no obvious reason why learning $(B - LD)$ should be more difficult than learning $B$.

The cancellation mechanism requires that the CNS learns to send out well-coordinated commands to gamma (for fusimotor control) and alpha motor neurons (for skeletomotor control). This requires additional learning next to learning the state observer. It is possible that the CNS initially relies on correction using efference copy information and, as learning progresses, gradually shifts to cancellation at the source. This would argue for the hybrid mechanism in Eq. 7.

There exists a third mechanism for cancelling reafferent feedback, and it depends on a so-called corollary discharge. This mechanism has been extensively documented and reviewed [74, 75]. A corollary discharge is an efference copy with a special role in the processing of sensory feedback: it cancels the reafferent feedback at the level of the CNS by preventing its input to the state observer. This differs from the mechanism described in this paper, in which all sensory feedback enters the state observer and regular efference copy information is used to cancel the reafferent component.

**Active perception**

Both the cancellation and the correction mechanism are a form of active perception. In the literature, there are different forms of active perception, and one familiar form refers to the fact that, to understand our environment, we move around to actively explore it [76, 77]. Here, I use "active perception" to denote that a percept (i.e., a state estimate) depends on both sensory information and information about the action (called "action information"); for cancellation, the action information is the fusimotor control signal, and for correction it is the efference copy. This is most clear for the scenario of a stationary joint. In this scenario, the cancellation mechanism produces a spindle firing rate that is proportional to the tension in the muscle. Because the tension depends on the joint angle (see Eq. 8), this allows the CNS to compute this joint angle. In turn, the correction mechanism takes as its sensory input a spindle firing rate that also depends on the action (via the reafferent acceleration $M^{-1}z$; see Eq. 6). However, via the efference copy, the CNS can correct for this action, allowing it to compute the joint angle.

**Incomplete sensory feedback in the vestibular system**

From the perspective of balance control, vestibular feedback would be complete if it informs the CNS about the body's state variables: tilt (roll and pitch) relative to gravity and the corresponding angular rates. Although the concept of incomplete feedback is absent in the vestibular system literature, much of the thinking is inspired by the fact that the output of the vestibular sensory organs (otoliths and semicircular canals) is only useful if it allows for balance control [78-82]. This awareness cannot be separated from state estimation because optimal state estimates are required for optimal control.

Otolith afferent fibers are sensitive to head tilt [78-80] via the gravitational force that acts on the endolymph fluid and thus in principle could provide feedback about the angular position of the body. However, every organ that is sensitive to gravitational force is also sensitive to forces that cause translational acceleration [78, 81, 82], and this results in the so-called gravito-inertial ambiguity. One way to resolve this ambiguity is by means of internal sensory model that maps the ambiguous otolith feedback on the state estimates. This is very much like the computational solution proposed in this paper: use an internal sensory model to map the ambiguous exafferent acceleration feedback (a linear combination of angular position and angular velocity) on the state estimates. Future research will have to show that this computational solution also allows for balance control based on vestibular feedback.

With respect to the semi-circular canals, the consensus is that the canal feedback is a high pass filtered angular velocity signal, with the filtering being due to the inertia of the endolymph fluid and its friction with the interior of the canals [73, 79, 83-89]. This feedback is incomplete but due to the high pass filtering, it cannot be expressed as a simple linear combination of the state variables. An important question is whether a computational solution exists for this type of incompleteness. A possible solution that has received a lot of attention in the literature is the so-called velocity buffer [79], which is an internal low pass filter of the expected angular velocity signal. This internal low pass filter can be added to the actual canal feedback, thereby possibly correcting for the high pass filtering. Future research will have to show whether this computational solution is also effective (i.e., allows for balance control).

An important difference between muscle spindle and vestibular feedback is that, for multi-joint mechanical systems, the latter is ambiguous with respect to the joint that is responsible for the feedback. This is because the sensory output of the otoliths and the canals depends on rotations over several possible axes: the ankle, the knee, the hip, and/or the neck. Future research will have to show whether a computational solution based on internal models allows for balance control using only this ambiguous vestibular feedback.

Just like muscle spindle feedback, vestibular feedback contains an exafferent and a reafferent component. There is very strong neurophysiological evidence for the fact that the CNS can suppress the reafferent feedback component. This evidence involves that, although the vestibular sensory afferents do not distinguish between active and passive movements (i.e., between reafferent and exafferent feedback), brainstem and cerebellar neurons show a reduced firing rate during active as compared to passive movements [see refs in 73, 90]. As for the mechanism behind this reafference cancellation, it has been shown that, during active movements, the firing rate did not simply reflect the exafferent component but was inhibited if the proprioceptive feedback of the self-motion could be predicted [91]. If there was a discrepancy between the predicted and actual proprioceptive consequences of self-motion, central vestibular neurons no longer preferentially encode vestibular exafference.

### More complex mechanical systems

My model for balance control is general in the sense that it only depends on the mechanical system's EoM and some parameters for which realistic values can be found. However, because the mechanical system may be complex, it may be difficult to obtain its EoM. For this reason, I considered a restricted form of cycling in which the rider keeps his legs still, does not use them to carry weight, and relies on a motor for propulsion. Under these restrictions, I could derive EoM for the rider-bicycle combination. However, most cyclists transfer a large part of their weight to the pedals and the handlebars; downhill mountain bikers (the balance artists in the cycling community) even do this for the full 100 percent. To model bicycle balance control with less constraints on the rider's movements, we need EoM for a more general mechanical system in which (1) the lower body is no longer a part of the rear frame, and (2) the AoS for the combined upper and lower body is formed by saddle, pedals, and handlebars. For the rest of the model (the sensory and the computational system), no new ingredients are needed.

## Implications for robotics

Joint acceleration feedback has been studied in robotics [92, 93] but following different approaches as in this paper. I have found no studies that use the Kalman filter to estimate the state from exafferent acceleration feedback or to correct the state estimate for reafferent feedback. In robotics, the equivalent of an efference copy can be obtained by splitting the controller output (e.g., a voltage signal to a motor) and sending one copy back to the controller. Future research will have to show whether such a controller based on the Kalman filter can solve important engineering problems.

## Conclusion

For a CNS that can perform state estimation, the debate in sensory neurophysiology about whether muscle spindles provide full state feedback is not crucial for motor control. Sensory models for both exafferent and total acceleration feedback allow for state estimation by the CNS. However, balance control using these state estimates is only possible if the reafferent acceleration component is either absent from the feedback or is corrected for.

# Methods

## Matlab toolbox

All simulations were performed using the Matlab Balance Control (BalCon) toolbox that is shared in the supplementary information, together with the scripts that produced the simulation results, including the figures. For every mechanical system, the BalCon toolbox contains one function that computes the EoM and their linearization. Most of the toolbox's functions are generic, in the sense that they can be used for all mechanical systems.

## Simulations of standing balance control

The computer simulations are based on difference equations that are discrete time versions of the differential equations that were presented in the Results section. The difference equations for the linear dynamical system under consideration are the following:

$$\begin{aligned} \boldsymbol{x}_{k+1} &= A_{dt}\boldsymbol{x}_k + B_{dt}\boldsymbol{u}_k + B_{dt}\boldsymbol{m}_k \\ \boldsymbol{y}_k &= C_{Comp}\boldsymbol{x}_k + D\boldsymbol{u}_k + B_{dt}^{lo}\boldsymbol{m}_k - B_{dt}^{lo}\boldsymbol{f}_k + \boldsymbol{s}_k \end{aligned}$$

In this equation, k indexes discrete time steps that are separated by a time interval *dt*. The matrices $A_{dt}$ and $B_{dt}$ are obtained from the well-known solution of a linear differential equation: $A_{dt} = e^{Adt}$ and $B_{dt} = A^{-1}(A_{dt} - I)B$, in which $I$ is the identity matrix [94]. The matrix $B_{dt}$ is the discrete time version of $B = M^{-1}$, and its lower half is denoted by $B_{dt}^{lo}$. The feedforward matrix $D$ is equal to the lower half of $B = M^{-1}$. Optimal control of this linear dynamical system is provided by the discrete time version of Eq. 18:

$$\widehat{\boldsymbol{x}}_{k+1} = A_{dt}\widehat{\boldsymbol{x}}_k + B_{dt}\boldsymbol{u}_k + L_{dt}\left[y_k - C_{Comp}(A_{dt}\widehat{\boldsymbol{x}}_k + B_{dt}\boldsymbol{u}_k) - D\boldsymbol{u}_k\right]$$

in which $\boldsymbol{u}_k = -K_{dt}(\hat{\boldsymbol{x}}_k - \boldsymbol{r})$. The matrices $L_{dt}$ and $-K_{dt}$ are, resp., the discrete time Kalman and LQR gain.

The mechanical system (the CIP) is specified by realistic and/or empirical values for its parameters: body length $l = 1{,}85$ m., body mass $m = 85$ kg., gravitational constant $g = 9{,}8066$, ankle stiffness $k_{stiff} = 493.4706$ Nm/rad. (64% of the critical stiffness), and ankle damping $c_{damp} = 30$ Nm/(rad./s) [42]. The CIP dynamics are simulated by the Matlab function ode45, which is based on an explicit Runge-Kutta (4,5) formula [95]. As depicted in Fig. 1, the input to the mechanical system is $\boldsymbol{z}_k = \boldsymbol{u}_k + \boldsymbol{m}_k$, with $\boldsymbol{u}_k = -K_{dt}(\hat{\boldsymbol{x}}_k - \boldsymbol{r})$, and the output is $\left[\boldsymbol{x}_{k+1}^t, \ddot{\boldsymbol{\theta}}_{k+1}\right]^t$.

The feedback loop is closed by the sensory system that maps the output of the mechanical system ($\ddot{\boldsymbol{\theta}}_{k+1}$), the output of the fusimotor system ($\boldsymbol{w}_{k+1}$), and the pure sensory noise ($\boldsymbol{s}_{k+1}$) to the sensory feedback $\boldsymbol{y}_{k+1}$:

$$\boldsymbol{y}_{k+1} = C_{Sens}\begin{bmatrix}\ddot{\boldsymbol{\theta}}_{k+1}\\ \boldsymbol{w}_{k+1}\end{bmatrix} + \boldsymbol{s}_{k+1}$$

To simulate CIP dynamics under closed-loop feedback control, one must add noise. I set the noise parameters such that the effects of motor and pure sensor noise on the sensory feedback are equal. As a common scale for the effects of these three noise types, I use the noise variance of the sensory feedback $\boldsymbol{y}_k$: if the noise source is only sensory, then $var(\boldsymbol{y}_k) = var(\boldsymbol{s}_k)$, and if the noise source is only motor, then

$$var(\boldsymbol{y}_k) = B_{dt}^{lo}\mathrm{var}(\boldsymbol{m}_k){B_{dt}^{lo}}^t + C_{Comp}B_{dt}\mathrm{var}(\boldsymbol{m}_k)B_{dt}^t C_{Comp}^t \quad \text{Eq. 19}$$

In the simulations, I start from a noise scaling variable $\sigma$ and (1) set $var(\boldsymbol{s}_k) = \sigma I$ ($I$ is an identity matrix of the appropriate size), (2) scale $var(\boldsymbol{m}_k)$ such that the expression on the right side of Eq. 19 equals $\sigma I$, and (3) set $var(\boldsymbol{f}_k) = var(\boldsymbol{m}_k)$. I set the covariance $cov(\boldsymbol{m}_k, \boldsymbol{f}_k)$ such that the shared variance of $\boldsymbol{m}_k$ and $\boldsymbol{f}_k$ is 50%.

The variance of the noise terms $B_{dt}\boldsymbol{m}_k$ (for the state equation, but also fed into the output equation) and $-B_{dt}^{lo}\boldsymbol{f}_k + \boldsymbol{s}_k$ (for the output equation only) and their covariance are used as weights in the calculation of the Kalman gain $L_{dt}$. This Kalman gain is a part of the computational system, and we thus implicitly assume that the CNS learns the sensor and the motor noise amplitudes from experience.

To be realistic, the ankle torque in the simulations must be less than the maximum voluntary contraction (MVC) of the ankle muscles. Therefore, the output of the computational system is truncated at an empirically determined MVC. Because humans prefer a leaned forward position, I use plantarflexion MVC. In a study with 20 participants, the mean plantarflexion MVC was estimated to be 195 Nm [53].

In the simulations, I evaluate whether the CoG remains over the AoS, which is a fraction of the sole length. To determine the AoS, I use a body-to-sole-length ratio of 6.6 [96], place the ankle at 25% from the rear end of the sole, and assume that the CoG must remain at 10% from the front and the rear end of the sole. The latter assumption is required because there can be no reaction forces near the edge of the sole. This results in a critical angular position interval [-0.0758, 0.2293] rad. over the ankle joint.

The LQR gain $-K_{dt}$ depends on the weights of the optimality criterium in Eq. 16, the 2-by-2 matrix $Q$ and the scalar $R$. These weights are set such that the precision ($Q$-dependent) and the energetic cost ($R$-dependent) component have an equal contribution to the optimality criterium. I use the maximum metric to calculate diagonal matrices $Q$ and $R$ that produce equal values for $\boldsymbol{x}_{max}^t Q \boldsymbol{x}_{max}$ and $\boldsymbol{u}_{max} R \boldsymbol{u}_{max}$. For the calculation of $\boldsymbol{x}_{max}$, I start from the critical angular position interval [-0.0758, 0.2293] rad., and calculate the critical angular velocity from the maximum angle position frequency of 1.875 Hz reported in [97]. This results in $\boldsymbol{x}_{max} = [0.2137 \quad 0.3655]^t$. I give angular position and angular velocity an equal contribution to the precision component by setting $Q_{11}$ and $Q_{22}$ such that $0.2137^2 Q_{11} = 0.3655^2 Q_{22}$. The critical torque $u_{max} = 195$ is obtained from [53].

## A general mechanical model constrains the relation between acceleration, force input, and state variables

I now demonstrate the relation between $C_{Sens}$ and $C_{Comp}$ for a general multibody mechanical system. I use this more general relation to model bicycle balance control using joint acceleration feedback from the rider's upper body. I start from the following differential equation for a multibody mechanical system:

$$M(\boldsymbol{\theta})\ddot{\boldsymbol{\theta}} + D(\boldsymbol{\theta}, \dot{\boldsymbol{\theta}}) = H\boldsymbol{z} \qquad \text{Eq. 20}$$

The vectors $\boldsymbol{\theta}$, $\dot{\boldsymbol{\theta}}$, $\ddot{\boldsymbol{\theta}}$ and $\boldsymbol{z}$ are, resp., the positions, velocities, accelerations and forcing torques of a multibody mechanical system. Because not all joints have to be actuated (see the bicycle model for an example), I use the matrix *H* to distribute the forcing torques over the joints; rows of *H* that correspond to non-actuated joints are zero. The matrix $M(\boldsymbol{\theta})$ is the mass moment of inertia and the vector $D(\boldsymbol{\theta}, \dot{\boldsymbol{\theta}})$ captures all forces that act on the passive system (gravity, centrifugal, damping, stiffness). In the CIP, the mass moment of inertia $M(\boldsymbol{\theta})$ is independent of $\boldsymbol{\theta}$ and was denoted by $M$. Note that Eq. 20 is more general than the familiar manipulator equation form [16], which splits $D(\boldsymbol{\theta}, \dot{\boldsymbol{\theta}})$ into two terms of which one depends on gravity only.

Eq. 20 can be rewritten as follows:

$$\ddot{\boldsymbol{\theta}} - M(\boldsymbol{\theta})^{-1}H\boldsymbol{z} = -M(\boldsymbol{\theta})^{-1}D(\boldsymbol{\theta}, \dot{\boldsymbol{\theta}}) \qquad \text{Eq. 21}$$

I now linearize the left side of Eq. 21 with respect to $\ddot{\boldsymbol{\theta}}$ and $\boldsymbol{\tau}$, and the right side with respect to $\boldsymbol{\theta}$ and $\dot{\boldsymbol{\theta}}$. Next, I evaluate the Jacobian of this linearization at the unstable fixed point $[\boldsymbol{\theta}; \dot{\boldsymbol{\theta}}] = [\mathbf{0}; \mathbf{0}]$. The Jacobian of the left side is $[I, -M(\mathbf{0})^{-1}H]$, in which $I$ is the identity matrix. The Jacobian of the right side requires symbolic differentiation, and the outcome of this operation is denoted by $J_{[\boldsymbol{\theta};\dot{\boldsymbol{\theta}}]}(\mathbf{0}; \mathbf{0})$. Inserting the linear approximations in the left- and the right side of Eq. 21, one obtains

$$[I, -M(\mathbf{0})^{-1}H] \begin{bmatrix} \ddot{\boldsymbol{\theta}} \\ \boldsymbol{z} \end{bmatrix} \approx J_{[\boldsymbol{\theta};\dot{\boldsymbol{\theta}}]}(\mathbf{0}; \mathbf{0}) \begin{bmatrix} \boldsymbol{\theta} \\ \dot{\boldsymbol{\theta}} \end{bmatrix} \qquad \text{Eq. 22}$$

The sensory feedback mapping matrices $C_{Sens}$ and $C_{Comp}$ are equal to the Jacobians in, resp., the left- and the right side of Eq. 22. We thus obtain

$$C_{Sens} \begin{bmatrix} \ddot{\boldsymbol{\theta}} \\ \boldsymbol{z} \end{bmatrix} \approx C_{Comp} \begin{bmatrix} \boldsymbol{\theta} \\ \dot{\boldsymbol{\theta}} \end{bmatrix}$$

Every row in $C_{Sens}$ specifies the difference between the acceleration at a single joint and a linear combination of the forcing torques. The corresponding row in $C_{Comp}$ specifies a linear combination of the state variables.

Note that these Jacobians also appear in the linearization of the EoM:

$$\ddot{\boldsymbol{\theta}} = -M(\boldsymbol{\theta})^{-1}D(\boldsymbol{\theta}, \dot{\boldsymbol{\theta}}) + M(\boldsymbol{\theta})^{-1}H\boldsymbol{z}$$

$$\ddot{\boldsymbol{\theta}} \approx J_{[\boldsymbol{\theta};\dot{\boldsymbol{\theta}}]}(\mathbf{0}; \mathbf{0}) \begin{bmatrix} \boldsymbol{\theta} \\ \dot{\boldsymbol{\theta}} \end{bmatrix} + M(\mathbf{0})^{-1}H\boldsymbol{z}$$

Thus, the EoM in state-space form, $[\dot{\boldsymbol{\theta}}; \ddot{\boldsymbol{\theta}}] = \dot{\boldsymbol{x}} = A\boldsymbol{x} + B\boldsymbol{z}$, have coefficient matrices $A$ and $B$, of which the lower halves are equal to, resp., $J_{[\boldsymbol{\theta};\dot{\boldsymbol{\theta}}]}(\mathbf{0}; \mathbf{0})$ and $M(\mathbf{0})^{-1}H$.

### The SDP EoM

The SDP EoM can be assembled from three components: (1) the EoM of a double compound pendulum with an actuated base (like the double compound pendulum on a cart, DCPC) and an actuated pelvic joint (like the Acrobot), (2) a planar kinematic bicycle model [98] that produces a formula for the acceleration at this base, and (3) a torsional spring-mass-damper system for the steering angle. Specifically, I model the angular positions $\theta_1$ and $\theta_2$ as the result of a double compound pendulum on a (zero-mass) cart (DCPC) and acceleration equal to $\alpha(\delta)$, the centrifugal acceleration derived under the planar kinematic bicycle model (see further). The EoM for this DCPC can be written as follows:

$$\begin{bmatrix} d_1 \cos(\theta_1) \\ d_2 \cos(\theta_2) \end{bmatrix} \alpha(\delta) + \begin{bmatrix} d_3 & d_4 \cos(\theta_1 - \theta_2) \\ d_4 \cos(\theta_1 - \theta_2) & d_5 \end{bmatrix} \begin{bmatrix} \ddot{\theta}_1 \\ \ddot{\theta}_2 \end{bmatrix} + \begin{bmatrix} 0 & d_4 \sin(\theta_1 - \theta_2)\, \theta_2 \\ d_4 \sin(\theta_1 - \theta_2)\, \phi_1 & 0 \end{bmatrix} \begin{bmatrix} \dot{\theta}_1 \\ \dot{\theta}_2 \end{bmatrix} + \begin{bmatrix} -f_1 \sin(\theta_1) \\ -f_2 \sin(\theta_2) \end{bmatrix} + \begin{bmatrix} k_{pelvis}(\theta_1 - \theta_2) + c_{pelvis}(\dot{\theta}_1 - \dot{\theta}_2) \\ -k_{pelvis}(\theta_1 - \theta_2) - c_{pelvis}(\dot{\theta}_1 - \dot{\theta}_2) \end{bmatrix} = \begin{bmatrix} 0 \\ T_\theta \end{bmatrix} \quad \text{Eq. 23}$$

These EoM are obtained by first applying the Euler-Lagrange method to the DCPC with a zero-mass cart, and then adding the constraint that the cart is controlled by the steering-induced centrifugal acceleration $\alpha(\delta)$. The derivation of the DCPC EoM using the Euler-Lagrange method can be found in the literature. Here, I started from Bogdanov (99) and added stiffness (with constant $k_{pelvis}$), damping (with constant $c_{pelvis}$) and torque input $T_\theta$ at the pelvic joint (between the upper and the lower body), similar to the Acrobot [16]. Next, I added the constraint that the angles $\theta_1$ and $\theta_2$ have no direct effect on the position of the base of the first rod (in the DCPC, the point where the cart is attached). This constraint follows from the fact that the bicycle's wheels are oriented perpendicular to the cart wheels of the DCPC. Under this constraint, the position of the base of the first rod is fully controlled by the steering-induced centrifugal acceleration $\alpha(\delta)$.

The constants in Eq. 23 are defined as follows:

$$d_1 = m_1 l_1 + m_2 L_1$$
$$d_2 = m_2 l_2$$
$$d_3 = m_1 {l_1}^2 + m_2 {L_1}^2 + I_1$$
$$d_4 = m_2 L_1 l_2 \qquad \text{Eq. 24}$$
$$d_5 = m_2 {l_2}^2 + I_2$$
$$f_1 = (m_1 l_1 + m_2 L_1) g$$
$$f_2 = m_2 l_2 g$$

The constants $m_1$, $L_1$, $l_1$ and $I_1$ are, resp., the mass, the length, the CoG ($L_1/2$) and the mass moment of inertia of the double pendulum's first rod, which represents the bicycle and the rider's lower body. The constants $m_2$, $L_2$, $l_2$ and $I_2$ are defined in the same way, but now for the second rod, which represents the rider's upper body. Finally, $g$ is the gravitational constant.

I now give the formula for the centrifugal acceleration $\alpha(\delta)$ that can be derived from a well-known planar bicycle model from the vehicle dynamics literature [98]:

$$\alpha(\delta) = v^2 \frac{\cos\big(\beta(\delta)\big)}{W} \tan(\delta)$$

This formula depends on the speed $v$, the bicycle wheelbase $W$, and the so-called slip angle $\beta(\delta)$, which is the angle between the velocity vector of the combined CoG and the LoS. This slip angle can be obtained as follows:

$$\beta(\delta) = \tan^{-1}\left(\frac{w_r \tan(\delta)}{W}\right)$$

In this equation, $w_r$ is the position of the combined CoG on the LoS. More precisely, $w_r$ is the distance between the road contact point of the rear wheel and the orthogonal projection of the combined CoG on the LoS. For realistic values ($W = 1.02$, $w_r = 0.3$, $-20^{\circ} < \delta < 20^{\circ}$), the slip angle $\beta(\delta)$ is almost a linear function of $\delta$:

$$\beta(\delta) \approx \frac{w_r \delta}{W}$$

For steering angles $-20^{o} < \delta < 20^{o}$, all deviations from linearity are less than 0.36%. In the simulations, I have used this approximation.

Finally, I introduce the model for the steering angle $\delta$. This model assumes that the steering angle is fully controlled by rider-applied forces on the handlebars; I thus ignore all forces that may contribute to a bicycle's self-stability. The steering assembly consists of the front wheel, the fork, the handlebars, and the rider's arms. I model this assembly as a torsional spring-mass-damper system:

$$M_{steer}\ddot{\delta} + c_{steer}\dot{\delta} + k_{steer}\delta = T_{\delta} \qquad \text{Eq. 25}$$

In this equation, $M_{steer}$ is the assembly's mass moment of inertia, $c_{steer}$ its damping, and $k_{steer}$ its stiffness. The input to the steering assembly is the net torque produced by the rider's arm muscles and denoted by $T_{\delta}$.

It is possible to derive expressions for the second derivatives $\ddot{\delta}$ and $\left[\ddot{\theta}_1, \ddot{\theta}_2\right]^T$ from Eq. 23 and Eq. 25. These expressions are complicated and not insightful. In my simulations, I use these expressions to define the state-space equations $\dot{\boldsymbol{x}} = \Omega(\boldsymbol{x}, \boldsymbol{u} + \boldsymbol{m})$ for the state variables $\boldsymbol{x} = \left[\delta, \theta_1, \theta_2, \dot{\delta}, \dot{\theta}_1, \dot{\theta}_2\right]^T$, external torques $\boldsymbol{u} = [T_{\delta}, T_{\theta}]^T$, and motor noise $\boldsymbol{m}$.

### The BDP EoM

The BDP is based on three ideas. The first idea is to follow the approach of [56] and derive linearized EoM for a bicycle with the rider's lower body rigidly attached to the rear frame and no upper body. These linearized EoM depend on a number of constants, and I chose these constants such that (1) the front frame is as similar as possible to the self-stable benchmark bicycle model described by [56], and (2) the lengths and masses are as similar as possible to the SDP. The second idea is to model the interactions between the upper body and the rear frame (which includes the lower body) by the linearized EoM of the double compound pendulum, similar to [100]. Finally, the third idea is to first derive the BDP EoM without stiffness and damping terms, and to add these terms only in the last step.

The approach of [56] involves a method to calculate the defining matrices of linearized EoM of the following type:

$$\boldsymbol{M}\begin{bmatrix}\ddot{\delta}\\ \ddot{\theta}_1\\ \ddot{\theta}_2\end{bmatrix}+\boldsymbol{C}\begin{bmatrix}\dot{\delta}\\ \dot{\theta}_1\\ \dot{\theta}_2\end{bmatrix}+\boldsymbol{K}\begin{bmatrix}\delta\\ \theta_1\\ \theta_2\end{bmatrix}=0$$

The matrices $\boldsymbol{M}$, $\boldsymbol{C}$ and $\boldsymbol{K}$ are functions of several constants (angles, lengths, masses, mass moments of inertia, gravitational acceleration, speed) that characterize the bicycle components and the internal forces that act on them. (I used the same symbols as [56], and therefore $\boldsymbol{M}$ has a different meaning as in the previous sections.) [56] only derived linearized EoM for bicycles with a rider that was rigidly attached to the rear frame. Thus, the upper body angular position $\theta_2$ is absent from their EoM. This missing component can be obtained by linearizing the double compound pendulum EoM which models the interactions between $\theta_1$ and $\theta_2$. Schematically, each of the matrices $\boldsymbol{M}$, $\boldsymbol{C}$ and $\boldsymbol{K}$ is composed as follows:

$$\begin{bmatrix}\mathrm{MP}(1,1) & \mathrm{MP}(1,2) & 0\\ \mathrm{MP}(2,1) & \mathrm{MP}(2,2) & 0\\ 0 & 0 & 0\end{bmatrix}+\begin{bmatrix}0 & 0 & 0\\ 0 & \mathrm{DP}(1,1) & \mathrm{DP}(1,2)\\ 0 & \mathrm{DP}(2,1) & \mathrm{DP}(2,2)\end{bmatrix}$$

in which "MP" denotes "Meijaard, Papadopoulos et al" [56], and "DP" denotes "Double Pendulum". The MP calculations were performed by means of the Matlab toolbox Jbike6 [101], in which I entered the constants for a bicycle with the rider's lower body rigidly attached to the rear frame and no upper body. This produced the constants $\mathrm{MP}(i,j)$ $(i,j=1,2)$ for $\boldsymbol{M}$, $\boldsymbol{C}$ and $\boldsymbol{K}$.

I now model the interactions between the upper body and the rear frame by the linearized EoM of the double compound pendulum. The nonlinear

EoM of the double compound pendulum are obtained from Eq. 23 by removing the terms that correspond to the centrifugal acceleration $\alpha(\delta)$, the stiffness and the damping:

$$\begin{bmatrix} d_3 & d_4\cos(\theta_1-\theta_2) \\ d_4\cos(\theta_1-\theta_2) & d_5 \end{bmatrix}\begin{bmatrix}\ddot{\theta}_1\\ \ddot{\theta}_2\end{bmatrix} + \begin{bmatrix} 0 & d_4\sin(\theta_1-\theta_2)\,\theta_2 \\ d_4\sin(\theta_1-\theta_2)\,\theta_1 & 0 \end{bmatrix}\begin{bmatrix}\dot{\theta}_1\\ \dot{\theta}_2\end{bmatrix} + \begin{bmatrix}-f_1\sin(\theta_1)\\ -f_2\sin(\theta_2)\end{bmatrix} = \begin{bmatrix}0\\ T_\theta\end{bmatrix}$$

I evaluate these EoM at $\theta_1 = \theta_2$ and replace $\sin(x)$ by its linear approximation near 0: $\sin(x) \approx x$. This results in

$$\begin{bmatrix} d_3 & d_4 \\ d_4 & d_5 \end{bmatrix}\begin{bmatrix}\ddot{\theta}_1\\ \ddot{\theta}_2\end{bmatrix} + \begin{bmatrix} -f_1 & 0 \\ 0 & -f_2 \end{bmatrix}\begin{bmatrix}\theta_1\\ \theta_2\end{bmatrix} = \begin{bmatrix}0\\ T_\theta\end{bmatrix}$$

The constants $d_3, d_4$ and $d_5$ contain elements that must be added to the matrix $\boldsymbol{M}$, and the constants $f_1$ and $f_2$ contain elements that must be added to the matrix $\boldsymbol{K}$ (for the definitions, see Eq. 24). I will use the notation $\mathrm{DP}(i,j)$ $(i,j = 1,2)$ to denote these elements. For $\boldsymbol{M}$, the following elements are added:

- $\mathrm{DP}(1,1) = m_2 {L_1}^2$
- $\mathrm{DP}(1,2) = \mathrm{DP}(2,1) = d_4 = m_2 L_1 l_2$
- $\mathrm{DP}(2,2) = d_5 = m_2 {l_2}^2 + I_2$

And for $\boldsymbol{K}$, the following elements are added:

- $\mathrm{DP}(1,1) = m_2 L_1 g$
- $\mathrm{DP}(2,2) = -f_2 = -m_2 l_2 g$

Finally, I added stiffness and damping terms that were also added to the SDP. The stiffness and damping terms were added to, resp., $\boldsymbol{K}$ and $\boldsymbol{C}$.

### Stiffness and damping parameters for the SDP and the BDP

Compared to the ankle joint, much less is known about the stiffness and damping of the steering and pelvic joint. These are not joints in the strict biomechanical sense because they involve more than the interface between two bones; the steering joint involves both the arms and part of the upper body, and the pelvic joint involves both the hip joint (head of femur and acetabulum) and the lumbosacral joint (lumbar spine and sacrum). For the steering joint, I calculated the stiffness from an empirically determined time constant, as described in [59]. And for the pelvic joint, I chose a stiffness coefficient such that the elastic force was 10 percent of the average (over upper and lower body) gravitational force; this allowed the upper and the lower body to fall with different accelerations. The damping ratio for both joints was set at 20, which is a strongly overdamped system. Strong damping was necessary for simulating the SDP; a critical damping ratio of 1 (the minimum damping that is required to suppress the spring-induced oscillations) resulted in numerical inaccuracies in the ODE solver (Matlab's ode45) even at very low noise levels. The BDP could be simulated for a wide range of damping ratio's, including critical damping.